\documentclass[11pt]{article}

\RequirePackage{amsthm,amsmath,amsfonts,amssymb}
\RequirePackage[numbers]{natbib}
\RequirePackage[colorlinks,citecolor=blue,urlcolor=blue]{hyperref}
\RequirePackage{graphicx}

\usepackage{subfigure}
\usepackage{ulem}

\newtheorem{theorem}{Theorem}[section]

\newtheorem{proposition}[theorem]{Proposition}
\newtheorem{corollary}[theorem]{Corollary}
\newtheorem{definition}[theorem]{Definition}

\newtheorem{remark}{Remark}

\begin{document}

\title{ Modeling and Decoupling Systemic Risk}

\author{Jingyu Ji$^{a,c}$,  Deyuan Li$^b$,  Zhengjun Zhang$^c$ \\
$^a$School of Data Science, Fudan University, Shanghai, China\\
$^b$School of Management, Fudan University, Shanghai, China\\
$^c$Department of Statistics, University of Wisconsin, Madison, WI, USA
}
\date{}
\maketitle

\begin{abstract}

Identifying systemic risk patterns in geopolitical, economic, financial, environmental, transportation, epidemiological systems and their impacts is the key to risk management. This paper proposes a new nonlinear time series model: autoregressive conditional accelerated Fr\'echet (AcAF) model and introduces two new endopathic and exopathic competing risk measures for better learning risk patterns, decoupling systemic risk, and making better risk management. The paper establishes the probabilistic properties of stationarity and ergodicity of the AcAF model. Simulation demonstrates the efficiency of the proposed estimators and the AcAF model's flexibility in modeling heterogeneous data. Empirical studies on the stock returns in S\&P 500 and the cryptocurrency trading show the superior performance of the proposed model in terms of the identified risk patterns, endopathic and exopathic competing risks, being informative with greater interpretability, enhancing the understanding of the systemic risks of a market and their causes, and making better risk management possible.

\end{abstract}

{\bf Keywords}:
{Time-varying tail risk},
{extreme value analysis},
{nonlinear time series},
{business statistics},
{systemic risk}.

\section{Introduction}
Systemic risk refers to the risk of collapse of an entire complex system due to the actions taken by the individual component entities or agents that comprise the system. Systemic risk may occur in almost every area, for example, financial crisis, flooding, forest fire, earthquake, market crash, economic crisis, global disease pandemic (like COVID-19), among many others (see \cite{zhang2020studying, Zhang2021rejoinder}). Typically, a system contains a number of risk sources, and once one comes first to collapse, the whole system is affected immediately, i.e., the risk sources are competing. When a disaster event (systemic risk) occurs, it may not be known what causes the event, i.e., its risk source. In such a scenario, it is of significance to decompose systemic risk into competing risks for learning risk patterns and better risk management.

Internal risk refers to the risk from shocks that are generated and amplified within the system. It stands in contrast to external risk, which relates to shocks that arrive from outside the system. Many systems (e.g., social, political, geopolitical, economic, financial, market, regional, global, environmental, transportation, epidemiological, material, chemical, and physical systems) are subject to both types of risk. For instance, the cargo ship MV Ever Given stuck in the Suez Canal on March 26, 2021, faced two major sources of risk. One is its internal operation errors corresponded to internal risk, and the other is strong winds and weather factors contributed to external risk. For more examples of risk decoupling, we refer the readers to \cite{danielsson2003endogenous}.

The occurrence of systemic risk is strongly correlated with extreme events. Modeling systemic risk through modeling extreme events is one of the essential topics in risk management. Many extreme events in history have been associated with systemic risk. Over the past two decades, extreme financial events have repeatedly shown their dramatic and adverse effects on the global economy, which include the Asian financial crisis in 1998, the subprime mortgage crisis of the United States in 2008, the European sovereign debt crisis in 2013, and the ``crash'' of Chinese stock market in 2015. Failure to recognize these extreme events' probability makes regulators and practitioners lack effective methods to deal with and prevent the financial crisis. As such, measuring and monitoring extreme financial events' risk is essential in financial risk management.

Extreme value theory (EVT) has been a powerful tool in risk analysis and is widely applied to model extreme events in finance, insurance, health, climate, and environmental studies (e.g., \cite{embrechts1999extreme}, \cite{mcneil2000estimation}, \cite{poon2004extreme}, \cite{deng2020statistical}). Extreme events often appeared dynamically and clustered in finance. In the literature,
\cite{smith2000bayesian}, \cite{bali2007conditional}, \cite{chavez2014extreme}, \cite{kelly2014tail}, \cite{massacci2017tail}, \cite{zhao2018modeling}, \cite{mao2018stochastic}, \cite{Koo2020}, and \cite{ji2021} investigate the overall dynamical tail risk structures.
In financial applications, \cite{chavez2016extreme} offer an extreme value theory-based statistical approach for modeling operational risk losses by taking into account dependence of the parameters on covariates and time; \cite{zhang2010estimation} study multivariate maxima of moving maxima (M4) processes and apply the methodology to model jumps in returns; \cite{daouia2018estimation} use the extreme expectiles to measure Value-at-Risk (VaR) and marginal expected shortfall;
\cite{harvey2013dynamic} studies the volatility clustering behavior which implies the extreme events' behavior and structure may also change as time goes by.


In the era of Big Data, data may come from multiple sources, and the data from each source has its own generating process, i.e., its probability distribution. The models mentioned above for overall tail risk cannot capture the sources of tail risk accurately. To model extreme values observed from different data sources, there exist some recent studies, e.g., \cite{heffernan2007asymptotically}, \cite{naveau2011extension}, \cite{tang2013sparse}, \cite{malinowski2015marked},  \cite{zhang2016copula} and \cite{idowu2017extended}. However, these models do not provide insights in risk sources, i.e., they do not differentiate different competing risks. Most recently, the accelerated max-stable distribution has been proposed by \cite{doi:10.1080/24754269.2020.1846115} to fit the extreme values of data generated from a mixture process (i.e., from different sources), whose mixture patterns vary with the time or sample size. The accelerated max-stable distributions form a new family of extreme value distributions for modeling maxima of maxima. They provide new probability foundations and statistical tools for modeling competing risks, e.g., endopathic and exopathic competing risks in this paper. The introduction of endopathic and exopathic competing risks are motivated from the widely used endogenous and exogenous variables in economic modeling. However, endopathic and exopathic competing risks in our model settings are tail-index implied and they show clear paths when clustered disaster events occur, and their interpretations as internal risks and external risks respectively are meaningful both quantitatively and qualitatively in a time series context.

This paper develops an endopathic and exopathic dynamic competing risks model that provides a new tool for better informative and rigorous tail risk analysis. The advantage of our model is to decouple systemic risk into endopathic and exopathic competing risks and measure them. Such decomposition methodology is new to the literature. Our model does not distinguish the data sources apriori, but refines the data's information, characterizes the dynamic tail risk behavior of extreme events through estimated parameter dynamics, and explicitly distinguishes the risk sources, i.e., endopathic risk and exopathic risk. 
The implementation uses autoregressive conditional accelerated Fr\'echet (AcAF) distributions to model systemic risks from different sources dynamically. The AcAF model can be applied to financial markets and many other areas where endopathic risk and exopathic risk are intertwined.

This paper makes the following contributions to the growing literature on tail risk measurement in the financial market and the literature in probability theory and time series, and many applied sciences. First and foremost, we propose a new decoupling risk framework to handle systemic risk. We decouple the systemic risk into endopathic risk and exopathic risk, which is the first based on our knowledge in the field. The AcAF model has two unique features. 1) Although we do not know which data sources the observations come from, the risks from different sources can be reconstructed through the estimated results. 2) The reconstructed parameter dynamics accurately capture the behavior of different risks. Second, the empirical analysis shows our model's superior performance in two financial markets: the U.S. stock market and the Bitcoin trading market. For the U.S. stock market, we find that exopathic risks are more volatile than endopathic risks. Under normal market conditions, endopathic risks dominate the stock market price fluctuations, while under turbulent market conditions, exopathic risks dominate. For the Bitcoin trading market, endopathic risks are more volatile than exopathic risks. Exopathic risks dominate the cryptocurrency market price fluctuations under normal market conditions, while under turbulent market conditions, endopathic risks dominate. The apparent opposite phenomena in these two markets are consistent with the actual market structure. Third, our technical proof is non-trivial and can not follow the existing literature's proof directly. They can be applied to other scenarios involving tail processes and parameter dynamics.

The rest of the paper is organized as follows. In Section \ref{sec: model}, we introduce the AcAF model and investigate its probabilistic properties. In Section \ref{sec: est}, we construct the conditional maximum likelihood estimators (cMLE) for estimation and provide a theory for the estimators' consistency and asymptotic normality. Simulation studies are presented in Section \ref{sec: sim}, which evaluate the performance of cMLE and the AcAF model's superior performance to the existing dynamic generalized extreme value (GEV) models for heterogeneous data. In Section \ref{sec: real}, we apply our model to three time series of maxima of maxima of negative log-returns in the stock market and Bitcoin market: one on the cross-sectional maximum losses (i.e., negative log-returns) of stocks in S\&P 500, one on the intra-day maximum losses of high-frequency trading of GE stock, and the other on the intra-day maximum losses of high-frequency Bitcoin trading. Section \ref{sec:con} gives concluding remarks and discussions. We conclude that the real data results show that our model has a strong ability to portray the endopathic and exopathic risks of the market and capture the market's dynamic endopathic and exopathic structure. All the technical details are given in the Appendix.
\section{Autoregressive conditional accelerated Fr\'echet model}\label{sec: model}
\subsection{Background and motivation}
In the era of Big Data, data generated from multiple sources meet in a commonplace. For instance, trading behavior in a market can be different from time to time, e.g., in the morning and the afternoon. Trading behavior in two different markets can be different at the same time. The recorded maximal signal strengths in a brain region can be dynamic, and their source origins can be different from time to time. The maximal precipitations/snowfalls/temperatures in a large area can be dynamic, and their exact locations can be different from time to time. In these examples, the available data are often in a summarized format, e.g., mean, median, low, high, i.e., not all details are given. As a result, the observed extreme values at a given time often come from different latent data sources with different populations. Certainly, the maxima resulted from each individual source has its data generating process, i.e., its limiting extreme value distribution is unique. As such, the classical extreme value theory cannot be directly applied to model the maxima drawn from different populations mixed together. The new EVT for maxima of maxima introduced by \cite{doi:10.1080/24754269.2020.1846115} provides the probabilistic foundation of accelerated max-stable distribution for studying extreme values of cross-sectional heterogeneous data. We will perform statistical modeling of extreme time series on the basis of this new EVT framework.

The autoregressive conditional Fr\'{e}chet (AcF) model in \cite{zhao2018modeling} portrays the time series of maxima well. Nevertheless, it does not directly model the heterogeneous data driven by two different risk factors, i.e., endopathic risk dynamics and exopathic risk dynamics. To further advance the new EVT of maxima of maxima and the AcF model, we propose the AcAF model to characterize different sources of tail risks in the financial market, under which a conditional evolution scheme is designed for the parameter $(\mu_t, \sigma_t, \alpha_{1t}, \alpha_{2t})^T$ of accelerated Fr\'echet distribution, so that time dependency and different risk sources of maxima of maxima can be captured.
\subsection{Model specification}
Suppose $Q_{kt}$, $k=1,...,d$ are latent processes, and $Q_t=\max_{1\leq k\leq d}Q_{kt}$ where each $Q_{kt}=\max_{1\leq i \leq p_{kt}}X_{k,i,t}$ is again maxima of many time series at time $t$. Following \cite{zhao2018modeling} and \cite{mao2018stochastic}, we assume
\begin{equation}
Q_{kt}=\mu_{kt}+\sigma_{kt}Y_{kt}^{1/\alpha_{kt}}\label{2.1},
\end{equation}
where $\mu_{kt}$, $\sigma_{kt}$ and $\alpha_{kt}$ are the location, scale, and shape parameters with $Y_{kt}$ being a unit Fr\'{e}chet random variable with the distribution function $F(y)=e^{-1/y},\ y>0$. Specifically, we consider two latent processes $Q_{1t}$ and $Q_{2t}$ to represent maximum negative log-returns across a group of stocks or of a particular stock's high-frequency trading whose price changes are driven by normal trading behavior and external information (e.g., sentiments), respectively. For example, with normal trading behavior the trading price changes of a particular stock can be higher during the market opening time and the market closing time, and with external information the trading price changes can be quite different from normal trading patterns, i.e., the price changes due to external information can occur at any time. The resulting maximum negative log-returns across that group of stocks or of that particular stock's high-frequency trading can be expressed as
$Q_t=\max(Q_{1t},Q_{2t})=\max(\max_{1 \leq i \leq p_{1t}}X_{1, i,t},\max_{1 \leq i \leq p_{2t}}X_{2,i,t})$, where each $\{X_{k,i,t}\}^{p_{kt}}_{i=1}$, $k=1, 2,$ is a set of time series whose price changes are due to corresponding price change driving factors, respectively. 

Note that $p_{1t}$ and $p_{2t}$ are the numbers of transactions, and they can be different and itself can be different from time to time, and the corresponding causes of price changes (negative log-returns) of $X_{1, i,t}$ and $X_{2,i,t}$ cannot be fully determined, i.e., $Q_{1t}$ and $Q_{2t}$ are unobservable latent processes. Here $Q_{1t}$ and $Q_{2t}$ do not correspond to the price changes during the market opening time and the market closing time which were used as a motivating example. They should be understood as they coexist all the time and the dominant one at any given time is observed.

For model parsimony, we assume $\mu_{1t}=\mu_{2t}=\mu_t$, $\sigma_{1t}=\sigma_{2t}=\sigma_t$, and follow the literature to assume $\mu_t$ as a constant and focus on the dynamics of $\sigma_t$, $\alpha_{1t}$ and $\alpha_{2t}$, which are the pivotal parameters of modeling systemic risk and identifying risk sources.
For the rest of the paper, we consider the following model:
\begin{eqnarray}
Q_t&=&\max(Q_{1t},Q_{2t})=\mu+\sigma_{t}\max(Y_{1t}^{1/\alpha_{1t}},Y_{2t}^{1/\alpha_{2t}}),\label{10}\\
\log\sigma_t&=&\beta_0+\beta_1 \log\sigma_{t-1}-\beta_2 \exp(-\beta_3 Q_{t-1}), \label{11}\\
\log\alpha_{1t}&=&\gamma_0+\gamma_1 \log\alpha_{1,(t-1)}+\gamma_2 \exp(-\gamma_3 Q_{t-1}), \label{12}\\
\log\alpha_{2t}&=&\delta_0+\delta_1 \log\alpha_{2,(t-1)}+\delta_2 \exp(-\delta_3 Q_{t-1}), \label{13}
\end{eqnarray}
where $\{Y_{1t}\}$ and $\{Y_{2t}\}$ are sequences of independent and identically distributed (i.i.d.) unit Fr\'{e}chet random variables. $Y_{1t}^{1/\alpha_{1t}}$ and $Y_{2t}^{1/\alpha_{2t}}$ can be considered as the normal trading driving factor and external information driving factor respectively as mentioned earlier. They compete against each other. The distribution of $\max(Y_{1t}^{1/\alpha_{1t}},Y_{2t}^{1/\alpha_{2t}})$ in equation (\ref{10}) is called the  accelerated Fr\'{e}chet distribution by \cite{doi:10.1080/24754269.2020.1846115}. In addition, $\beta_0, \gamma_0, \delta_0, \mu \in \mathbb{R}$, $0 \leq \beta_1 \neq \gamma_1 \neq \delta_1 <1$ and $\beta_2,\beta_3,\gamma_2,\gamma_3,\delta_2,\delta_3>0$ are assumed for the model to be stationary and technical requirements.
\begin{remark}\label{simplified}
	Note that although $\beta_2$, $\gamma_2$, $\delta_2$ are all assumed greater than zero, they can be set as zero. As long as any of them are set to be zero, all theories and estimation methods developed can be easily adjusted because the corresponding dynamic equations will become constants. For example, assuming  $\gamma_2=0$, then the dynamic equation (\ref{12}) will result in a stationary solution of $\alpha_{1t}=\exp(\gamma)$, where $\gamma=\gamma_0/(1-\gamma_1)$. This paper will not separately develop additional theoretical results for any of $\beta_2$, $\gamma_2$, and $\delta_2$ being zero as we will have a simplified model with the corresponding $\alpha_{1t}$ as a constant in Section \ref{section5.2}.
\end{remark}

Note that $\log\alpha_{1t}$ and $\log\alpha_{2t}$ share the same dynamic structure in (\ref{12}) and (\ref{13}). The following remarks solve the identifiability problem and give unique solution in the estimation.
\begin{remark}\label{alphaidentifiability}
	When one of $\gamma_2$ and $\delta_2$ is zero, we set $\alpha_{1t}$ as a constant equal to $\exp(\gamma)$ for model identifiability, where $\gamma=\gamma_0/(1-\gamma_1)$ or $\delta_0/(1-\delta_1)$, depending on which of $\gamma_2$ and $\delta_2$ is zero. When both $\gamma_2$ and $\delta_2$ are zero, we set $\alpha_{1t}$ corresponds to the smaller one. When both  $\gamma_2$ and $\delta_2$ are greater than zero, for model identifiability and under the stationary and ergodic properties of the $\{Q_t\}$ process, we assume $var(\gamma_2 \exp(-\gamma_3 Q_{t}))>var(\delta_2 \exp(-\delta_3 Q_{t}))$, which will be determined by the sample variances in real data applications.
\end{remark}

\begin{remark}\label{remark3}
	When both $\gamma_1$ and $\gamma_3$ (or $\delta_1$ and $\delta_3$) are zero, we set $\alpha_{1t}$ as a constant equal to $\exp(\gamma')$ for model identifiability, where $\gamma'=\gamma_0+\gamma_2$ or $\delta_0+\delta_2$, depending on which group of $\gamma_1$ and $\gamma_3$ or $\delta_1$ and $\delta_3$ are zero. Like in Remark \ref{simplified}, we have a simplified model with the corresponding $\alpha_{1t}$ as a constant in Section \ref{section5.2}.
\end{remark}

We now introduce our proposed endopathic risk and exopathic risk.

\begin{definition}\label{riskdecoup}
	When one of $\gamma_2$ and $\delta_2$ is zero, we refer $\alpha_{1t}$ to the tail index implied endopathic risk (for simplicity, call it endopathic risk) and $\alpha_{2t}$ to the tail index implied exopathic risk (for simplicity, call it exopathic risk). When both $\gamma_2$ and $\delta_2$ are zero, we define $\alpha_{1t}$ as the endopathic risk, while the exopathic risk is not defined. When both  $\gamma_2$ and $\delta_2$ are greater than zero, we refer $\alpha_{1t}$ to the endopathic risk and $\alpha_{2t}$ to the exopathic risk. When $\gamma_1$ and $\gamma_3$ (or $\delta_1$ and $\delta_3$) are zero, we define  $\alpha_{1t}$ as the endopathic risk and $\alpha_{2t}$ as the exopathic risk.
\end{definition}

The following two remarks rationalize the validity of the definitions of these two new endopathic risk and exopathic risk.

\begin{remark}
	When $\{\alpha_{1t}\}$ is a constant over time, it means that the corresponding tail index information is inherently built in $\{Q_{t}\}$ and is not varying with the time (even during clustered extreme events such as financial crisis due to external risks), i.e., it corresponds to internal risk or endopathic risk.
\end{remark}
\begin{remark}
	For both  $\gamma_2$ and $\delta_2$ being greater than zero, we define endopathic risks and exopathic risks using the idea in defining endogeneous variables and exogeneous variables in the economic modeling. It can be shown that under the correctly specified model (e.g., a regression model), the variance of the error term in a model with all exogenous variables included is the smallest compared to the variances of error terms in incorrectly specified models (e.g., missing covariates). By assuming $var(\gamma_2 \exp(-\gamma_3 Q_{t}))>var(\delta_2 \exp(-\delta_3 Q_{t}))$, we define $\alpha_{2t}$ to be the exopathic risks, and $\alpha_{1t}$ as the endopathic risks. Note that endopathic risks and exopathic risks are defined for time series with clustered extreme events, and their interpretations are very different from endogeneous variables and exogeneous variables. 
\end{remark}

In real data section, we use three examples to empirically justify the validity of the definitions.

\begin{remark}\label{How many alpha}
	In the model (\ref{10})-(\ref{13}), we set $\alpha_{it}$, $i=1,2$. A natural question will be why not make $i=1,2,3,...,k$ with $k>2$. Of course, making $k>2$ can be done theoretically and the probabilistic properties of the model can still hold. However, $k>2$ will increase statistical inference complexity and estimation inefficiency, e.g., in optimization problems. In the economic modeling literature, risks are often decoupled into two main risks, i.e., endogenous (internal) and exogenous (external), for easy interpretability. Certainly, internal (external) risks can further be decoupled into more specific risks, which can be challenging. Following the economic literature, we set $k=2$ in this paper.
\end{remark}

We note that the autoregressive structures used in $\sigma_t$, $\alpha_{1t}$ and $\alpha_{2t}$ can be traced back to GARCH model in \cite{bollerslev1986generalized}, autoregressive conditional density model in \cite{hansen1994autoregressive}, and autoregressive conditional durational model in \cite{engle1998autoregressive}. The clustering of extreme events in time is a significant feature of the extreme value series $\{Q_t\}$ in many applications, especially in financial time series. Empirical evidences have shown that extreme observations tend to happen around the same period in many applications. Translating this phenomenon in our model, we can say that an extreme event observed at time $t-1$ causes the distribution of $Q_t$ to have larger scale (large $\sigma_t$) and heavier tail (small tail index), resulting in a larger tail risk of $Q_t$. Here, a smaller tail index implies a larger tail risk. In Section \ref{example}, we present a class of factor models and show the limiting distribution of maxima of maxima of the response variables to be the accelerated Fr\'{e}chet types.  We next prove the stationarity and ergodicity of the AcAF model.
\subsection{Stationarity and ergodicity}\label{se}
The evolution schemes \eqref{11}-\eqref{13} can be written as
\begin{eqnarray}
\log\sigma_t&=&\beta_0+\beta_1 \log\sigma_{t-1}-\beta_2 \exp[-\beta_3 (\mu+\sigma_{t-1} \max(Y_{1,t-1}^{1/\alpha_{1,t-1}}, Y_{2,t-1}^{1/\alpha_{2,t-1}}))], \label{17}\\
\log\alpha_{1t}&=&\gamma_0+\gamma_1 \log\alpha_{1,t-1}+\gamma_2 \exp[-\gamma_3  (\mu+\sigma_{t-1} \max(Y_{1,t-1}^{1/\alpha_{1,t-1}}, Y_{2,t-1}^{1/\alpha_{2,t-1}}))], \label{18}\\
\log\alpha_{2t}&=&\delta_0+\delta_1 \log\alpha_{2,t-1}+\delta_2 \exp[-\delta_3  (\mu+\sigma_{t-1} \max(Y_{1,t-1}^{1/\alpha_{1,t-1}}, Y_{2,t-1}^{1/\alpha_{2,t-1}}))]. \label{19}
\end{eqnarray}
Hence $\{\sigma_t, \alpha_{1t}, \alpha_{2t}\}$ forms a homogeneous Markov chain in $\mathbb{R}^3$. The following theorem provides a  sufficient condition under which the process $\{\sigma_t, \alpha_{1t}, \alpha_{2t}\}$ is stationary and ergodic.
\begin{theorem}\label{theorem1}
	For the AcAF model with $\beta_0, \gamma_0, \delta_0,\mu \in \mathbb{R}$, $\beta_2, \beta_3, \gamma_2, \gamma_3, \delta_2, \delta_3>0$, and $0 \leq \beta_1 \neq \gamma_1 \neq \delta_1 <1$, the latent process $\{\sigma_t, \alpha_{1t}, \alpha_{2t}\}$ is stationary and geometrically ergodic.
\end{theorem}

The proof of Theorem \ref{theorem1} can be found in the Appendix. Since $\{Q_t\}$ is a coupled process of $\{\sigma_t, \alpha_{1t}, \alpha_{2t}\}$ through (\ref{10}), $\{Q_t\}$ is also stationary and ergodic.
\subsection{AcAF model under a factor model setting}\label{example}
In this section, we show that the limiting distribution of maxima $Q_t$ under a factor model framework coincides with the distribution of an AcAF model. We assume both $\{X_{1,i,t}\}_{i=1}^{p_{1t}}$ and $\{X_{2,j,t}\}_{j=1}^{p_{2t}}$ follow general factor models,
\begin{equation*}
\begin{aligned}
X_{1,i,t}&=f_i(Z_{1t}, Z_{2t}, \cdots, Z_{dt})+\sigma_{it} \epsilon_{1,i,t}, \\
X_{2,j,t}&=\tilde{f}_j(Z_{1t}, Z_{2t}, \cdots, Z_{dt})+\tilde{\sigma}_{jt} \epsilon_{2,j,t}, \\
\end{aligned}
\end{equation*}
where $\{X_{1,i,t}\}_{i=1}^{p_{1t}}$ and $\{X_{2,j,t}\}_{j=1}^{p_{2t}}$ are two latent time series at time $t$, $\{Z_{1t}, Z_{2t}, \cdots, Z_{dt}\}$ consist of observed and unobserved factors, $\{\epsilon_{1,i,t}\}_{i=1}^{p_{1t}}$ and $\{\epsilon_{2,j,t}\}_{j=1}^{p_{2t}}$ are two i.i.d. random noises that are independent to each other and independent with the factors $\{Z_{it}\}_{i=1}^d$, and $\{\sigma_{it}\}_{i=1}^{p_{1t}},  \{\tilde{\sigma}_{jt}\}_{j=1}^{p_{2t}}$ $\in \mathcal{F}_{t-1}$ are the conditional volatilities of $\{X_{1,i,t}\}_{i=1}^{p_{1t}}$ and $\{X_{2,j,t}\}_{j=1}^{p_{2t}}$, respectively. The functions $f_i, \tilde{f}_j : \mathbb{R}^d \to \mathbb{R}$ are Borel functions. Without misunderstanding, we use $p_1$ and $p_2$ to denote $p_{1t}$ and $p_{2t}$, respectively.

One fundamental characteristic of many financial time series is that they are often heavy-tailed. To incorporate this observation, we make the common assumption that the random noises $\{\epsilon_{1,i,t}\}_{i=1}^{p_1}$ and $\{\epsilon_{2,j,t}\}_{j=1}^{p_2}$ are i.i.d. random variables in the Domain of Attraction of Fr\'echet distribution (\cite{leadbetter2012extremes}). Here and after, for two positive functions $m_1(x)$ and $m_2(x)$, $m_1(x) \sim m_2(x)$ means $\frac{m_1(x)}{m_2(x)}\to 1$, as $x \to \infty$. Specifically, we adopt the following definition.

\begin{definition}[Domain of Attraction of Fr\'echet distribution]\label{def1}
	A random variable $\epsilon$ is in the Domain of Attraction of Fr\'echet distribution with tail index $\alpha$ if and only if $x_F=\infty$ and $1-F_{\epsilon} \sim l(x)x^{-\alpha}$, $\alpha>0$, where $F_{\epsilon}$ is the cumulative distribution function (c.d.f.) of $\epsilon$, $l(x)$ is a slowly-varying function and $x_F=\sup\{x: F_{\epsilon}(x)<1\}$.
\end{definition}

Domain of Attraction of Fr\'echet distribution includes a broad class of distributions such as Cauchy, Burr, Pareto and $t$ distributions. To facilitate algebraic derivation, we further assume that for slowly varying functions corresponding to $\{\epsilon_{1,i,t}\}_{i=1}^{p_1}$ and $\{\epsilon_{2,j,t}\}_{j=1}^{p_2}$ respectively, $l_{1t}(x) \to K_{1t}$ and $l_{2t}(x) \to K_{2t}$ as $x \to \infty$, where  $K_{1t}, K_{2t} \in \mathcal{F}_{t-1}$ are two positive constants. This is a rather weak assumption with all the aforementioned distributions satisfying this condition. Since $K_{1t}$ and $K_{2t}$ can be incorporated into each $\sigma_{it}$, without loss of generality, we set $K_{1t}$ and $K_{2t}$ are both equal to 1 in the following. Under a dynamic model, we assume that the conditional tail indices $\alpha_{1t}$ and $\alpha_{2t}$ of $\epsilon_{1,i,t}$ and $\epsilon_{2,j,t}$ respectively evolve through time according to certain dynamics (e.g., (\ref{12}) and (\ref{13})) and $\alpha_{1t}$, $\alpha_{2t}$ $\in$ $\mathcal{F}_{t-1}$.

We also assume that
\begin{equation*}
\sup_{1 \leq p_1 < \infty} \sup_{1 \leq i \leq p_1} |f_i(Z_{1t}, Z_{2t}, \cdots, Z_{dt})| < \infty,~\text{a.s.}
\end{equation*}
and that
\begin{equation*}
\sup_{1 \leq p_2 < \infty} \sup_{1 \leq j \leq p_2} |\tilde{f}_j(Z_{1t}, Z_{2t}, \cdots, Z_{dt})| < \infty, ~\text{a.s.}
\end{equation*}
Notice here the supremum is taken over $p_1$ or $p_2$ with the number of latent factors $d$ fixed. This is a  mild assumption and it includes all the commonly encountered factor models. For example, if the factor model takes a linear form, $f_i(Z_{1t}, \cdots, Z_{dt})=\sum_{s=1}^d \beta_s^{(i)}Z_{st}$, a sufficient condition for the assumption to hold would be $\sup_{1 \leq p_1 < \infty} \sup_{1 \leq i \leq p_1} \| \boldsymbol{\beta}^{(i)}  \| < \infty$, where $\boldsymbol{\beta}^{(i)}=(\beta_1^{(i)}, \cdots, \beta_d^{(i)})^T$. We further assume that there exist positive constants $C_1$ and $C_2$ such that $C_1 \leq \sigma_{it}, \tilde{\sigma}_{jt} \leq C_2$ for any $p_1, p_2$, $1 \leq i \leq p_1$ and $1 \leq j \leq p_2$.

Based on Proposition 1 in \cite{zhao2018modeling}, given $\mathcal{F}_{t-1}$, we have, as $p_1 \to \infty$, $p_2 \to \infty$,
\begin{equation*}
\frac{\max_{1 \leq i \leq p_1} \{X_{1,i,t}\} - b_{1,p_1,t}}{a_{1,p_1,t}} \stackrel{d}{\rightarrow} \Psi_{\alpha_{1t}} ~ \text{and}~~~
\frac{\max_{1 \leq j \leq p_2} \{X_{2,j,t}\} - b_{2,p_2,t}}{a_{2,p_2,t}} \stackrel{d}{\rightarrow} \Psi_{\alpha_{2t}},
\end{equation*}
where $a_{1,p_1,t}=(\sum_{i=1}^{p_1} \sigma_{it}^{\alpha_{1t}} )^{1/\alpha_{1t}}$, $a_{2,p_2,t}=( \sum_{j=1}^{p_2} \tilde{\sigma}_{jt}^{\alpha_{2t}} )^{1/\alpha_{2t}}$,  $b_{1,p_1,t}=b_{2,p_2,t}=0$, $\Psi_{\alpha_{1t}}(x)=\exp(-x^{-\alpha_{1t}})$ and $\Psi_{\alpha_{2t}}(x)=\exp(-x^{-\alpha_{2t}})$ denote the distributions of Fr\'echet type random variables with tail indices $\alpha_{1t}>0$  and $\alpha_{2t}>0$, respectively.

Recall that $Q_t=\max(Q_{1t}, Q_{2t})=\max\left(\max_{1 \leq i \leq p_1} X_{1,i,t},  \max_{1 \leq j \leq p_2} X_{2,j,t} \right)$. 
The limiting distribution form of $Q_t$ needs some discussions about the size of two tail indices and the order of $p_1$ and $p_2$.

\begin{proposition}\label{prop3}
	Given $\mathcal{F}_{t-1}$, under the assumptions in this section, the limiting distribution of $Q_t$ as $p_1, p_2 \to \infty$ can be determined in the following cases:
	\begin{description}
		\item[Case 1.] $\alpha_{1t}<\alpha_{2t}$.
		\begin{itemize}
			\item[1.] \noindent If $p_1/p_2\to C>0~\text{or}~\infty$, then $a_{1,p_1,t}/a_{2,p_2,t} \to \infty$ and
			$P\left(\frac{Q_t-b_{1,p_1,t}}{a_{1,p_1,t}} \leq x \right)   \to  \Psi_{\alpha_{1t}}(x).$
			\item[2.]  \noindent If $p_1/p_2 \to 0$ and $a_{1,p_1,t}/a_{2,p_2,t} \to a_t>0$, then
			$	P\left(\frac{Q_t-b_{1,p_1,t}}{a_{1,p_1,t}} \leq x \right)   \to \Psi_{\alpha_{1t}}(x) \Psi_{\alpha_{2t}}(a_tx).$
		\end{itemize}
		\item[Case 2.]
		$\alpha_{1t}=\alpha_{2t}=\alpha_t$.
		\begin{itemize}
			\item[1.]  \noindent If $p_1/p_2\to C>0$ and $a_{1,p_1,t}/a_{2,p_2,t} \to a_t>0$, then
			$P\left(\frac{Q_t-b_{1,p_1,t}}{a_{1,p_1,t}} \leq x \right)   \to \Psi_{\alpha_t}(x) \Psi_{\alpha_t}(a_tx).$
			\item[2.] \noindent If $p_1/p_2\to 0$, then $a_{1,p_1,t}/a_{2,p_2,t} \to 0$ and
			$	P\left(\frac{Q_t-b_{2,p_2,t}}{a_{2,p_2,t}} \leq x \right)  \to \Psi_{\alpha_t}(x).$
			\item[3.] \noindent If $p_1/p_2\to \infty$, then $a_{1,p_1,t}/a_{2,p_2,t} \to \infty$ and
			$P\left(\frac{Q_t-b_{1,p_1,t}}{a_{1,p_1,t}} \leq x \right)   \to \Psi_{\alpha_t}(x).$
		\end{itemize}
		\item[Case 3.]
		$\alpha_{1t}>\alpha_{2t}$.
		\begin{itemize}
			\item[1.]  \noindent If $p_1/p_2 \to C\geq0$, then $a_{1,p_1,t}/a_{2,p_2,t} \to 0$ and
			$P\left(\frac{Q_t-b_{2,p_2,t}}{a_{2,p_2,t}} \leq x \right) \to  \Psi_{\alpha_{2t}}(x).$
			\item[2] \noindent   If $p_1/p_2 \to \infty$ and $a_{2,p_2,t}/a_{1,p_1,t} \to a_t >0$, then
			$P\left(\frac{Q_t-b_{2,p_2,t}}{a_{2,p_2,t}} \leq x \right)   \to \Psi_{\alpha_{1t}}(a_tx) \Psi_{\alpha_{2t}}(x).$
		\end{itemize}
	\end{description}
\end{proposition}

The proof of Proposition \ref{prop3} can be found in the Appendix.

Under a particular setup, we assume $p_1=p_2=p$ and denote $f_i(Z_{1t}, Z_{2t}, \cdots, Z_{dt})=\tilde{f}_i(Z_{1t},$ $ Z_{2t}, \cdots, Z_{dt})$ as the underlying return values of the $i$th stock, $X_{1,i,t}$ (or $X_{2,i,t}$) as the unobserved value of the $i$th stock when the endopathic shock is stronger (weaker) than the exopathic shock. Under this setting, we can rewrite the observed time series $Q_t$ as
\begin{eqnarray}
Q_{t}&=&\max(X_{1t}, X_{2t})= \max\Big(\max_{1 \leq i \leq p}X_{1,i,t},\max_{1 \leq i \leq p}X_{2,i,t}\Big)\nonumber\\
&=&\max \left(\max_{1 \leq i \leq p} \Big(f_i(Z_{1t}, Z_{2t}, \cdots, Z_{dt}) + \sigma_{it} \epsilon_{1,i,t} \Big) , \max_{1 \leq i \leq p} \Big(f_i(Z_{1t}, Z_{2t}, \cdots, Z_{dt}) + \sigma_{it} \epsilon_{2,i,t} \Big)  \right)\nonumber\\
&=&\max_{1 \leq i \leq p} \Big(\max\Big( f_i(Z_{1t}, Z_{2t}, \cdots, Z_{dt}) + \sigma_{it} \epsilon_{1,i,t},  f_i(Z_{1t}, Z_{2t}, \cdots, Z_{dt}) + \sigma_{it} \epsilon_{2,i,t}    \Big)   \Big)\nonumber\\
&=& \max_{1 \leq i \leq p} \Big( f_i(Z_{1t}, Z_{2t}, \cdots, Z_{dt}) +\sigma_{it} \max(\epsilon_{1,i,t}, \epsilon_{2,i,t}) \Big).\nonumber
\end{eqnarray}

Corollary \ref{coro1} gives the general asymptotic conditional distribution of maxima $Q_t$ when $p$ goes to infinity.
\begin{corollary}\label{coro1}
	Denote $a_{j,p,t}=\left( \sum_{i=1}^p \sigma_{it}^{\alpha_{jt}} \right)^{1/\alpha_{jt}}$, and $b_{j,p,t}=0$ for $j=1,2$. Given $\mathcal{F}_{t-1}$, the limiting distribution of $Q_t$ as $p \to \infty$ can be determined in the following cases:
	\begin{itemize}
		\item[1.] If $\alpha_{1t}<\alpha_{2t}$, then $a_{1,p,t}/a_{2,p,t} \to \infty$, and
		$P \left( \frac{Q_t-b_{1,p,t} }{a_{1,p,t}} \leq x   \right)  \to \Psi_{\alpha_{1t}}(x).$ 
		\item[2.] If $\alpha_{1t}=\alpha_{2t}=\alpha_t$, then $a_{1,p,t}=a_{2,p,t}$, and
		$P\left( \frac{Q_t-b_{1,p,t} }{a_{1,p,t}} \leq x   \right)  \to  \Psi_{\alpha_t}(x) \Psi_{\alpha_t}(x).$ 
		\item[3.] If $\alpha_{1t}>\alpha_{2t}$, then $a_{1,p,t}/a_{2,p,t} \to 0$, and
		$P \left( \frac{Q_t-b_{2,p,t} }{a_{2,p,t}} \leq x   \right) \to \Psi_{\alpha_{2t}}(x).$ 
	\end{itemize}
\end{corollary}

Both Proposition \ref{prop3} and Corollary \ref{coro1} show that under the framework of the general factor model and some mild conditions, the conditional distribution of maxima $Q_t$ can be well approximated by an accelerate Fr\'echet distribution. In terms of stochastic representation, the observed maxima value $Q_t$ can be rewritten as $Q_t \approx \sigma_t \max(Y_{1t}^{1/\alpha_{1t}}, Y_{2t}^{1/\alpha_{2t}})$, where $Y_{1t}$ and $Y_{2t}$ are two independent unit Fr\'echet random variables and $\sigma_t$ depends on the size of $\alpha_{1t}$ and $\alpha_{2t}$. More specifically, if $\alpha_{1t}<\alpha_{2t}$, then $\sigma_{t}=\lim_{p \to \infty} a_{1,p,t}$; if $\alpha_{1t}>\alpha_{2t}$, then $\sigma_{t}=\lim_{p \to \infty} a_{2,p,t}$; if $\alpha_{1t}=\alpha_{2t}$, then $\sigma_{t}=\lim_{p \to \infty} a_{1,p,t}=\lim_{p \to \infty} a_{2,p,t}$. To be more flexible and accurate in finite samples, a location parameter $\mu_t$ can be included. That is,
$$Q_t \approx \mu_t +\sigma_t \max(Y_{1t}^{1/\alpha_{1t}},  Y_{2t}^{1/\alpha_{2t} }),$$
where $\{\mu_t, \sigma_t, \alpha_{1t}, \alpha_{2t}\}$ are time-varying parameters. Setting $\mu_t=\mu$ for parsimonious modeling, we obtain the dynamic structure of $\{Q_t\}$ specified in (\ref{10}).

\section{Parameter estimation}\label{sec: est}
We denote all the parameters in the model by $$\boldsymbol{\theta}=(\beta_0, \beta_1, \beta_2, \beta_3, \gamma_0, \gamma_1, \gamma_2, \gamma_3, \delta_0, \delta_1, \delta_2, \delta_3, \mu)^T,$$ and denote the parameter space by $$\Theta_s=\{\boldsymbol{\theta}\vert \beta_0, \gamma_0, \delta_0, \mu \in \mathbb{R}, 0 \leq \beta_1, \gamma_1, \delta_1 \leq 1, \beta_2, \beta_3, \gamma_2, \gamma_3, \delta_2, \delta_3 >0\}.$$ In the following, we assume that all allowable parameters are in $\Theta_s$ and the true parameter is $\boldsymbol{\theta}_0=(\beta_0^0, \beta_1^0, \beta_2^0, \beta_3^0, \gamma_0^0, \gamma_1^0, \gamma_2^0, \gamma_3^0, \delta_0^0, \delta_1^0, \delta_2^0, \delta_3^0, \mu_0)^T$.

The conditional probability density function (p.d.f.) of $Q_t$ given $(\mu,\sigma_t,\alpha_{1t},\alpha_{2t})^T$ is
\begin{equation}
\begin{aligned}
f_t(\boldsymbol{\theta})&=\{\alpha_{1t} \sigma_t^{\alpha_{1t}}(Q_t-\mu)^{-\alpha_{1t}-1}+ \alpha_{2t} \sigma_t^{\alpha_{2t}}  (Q_t-\mu)^{-\alpha_{2t}-1}\}\\
&~\quad \times  \exp\{-\sigma_t^{\alpha_{1t}} (Q_t-\mu)^{-\alpha_{1t}}-\sigma_t^{\alpha_{2t}}  (Q_t-\mu)^{-\alpha_{2t}}\}.
\end{aligned}
\end{equation}

By conditional independence, the log-likelihood function with observations $\{Q_t\}_{t=1}^n$ is
\begin{equation}
\begin{aligned}
L_n(\boldsymbol{\theta})&=\frac{1}{n}\sum_{t=1}^n l_t(\boldsymbol{\theta})=\frac{1}{n}\sum_{t=1}^n \bigg[\log \big\{\alpha_{1t} \sigma_t^{\alpha_{1t}} (Q_t-\mu)^{-\alpha_{1t}-1} +\alpha_{2t} \sigma_t^{\alpha_{2t}} (Q_t-\mu)^{-\alpha_{2t}-1} \big\}\\
&\quad-\sigma_t^{\alpha_{1t}}(Q_t-\mu)^{-\alpha_{1t}}-\sigma_t^{\alpha_{2t}}(Q_t-\mu)^{-\alpha_{2t}}\bigg],
\end{aligned}
\end{equation}
where $\{\sigma_t,\alpha_{1t},\alpha_{2t}\}_{t=1}^n$ can be obtained recursively through \eqref{11}-\eqref{13}, with an initial value $(\sigma_1,\alpha_{11},\alpha_{21})^T$.

Denote the log-likelihood function based on an arbitrary initial value $(\tilde{\sigma}_1, \tilde{\alpha}_{11}, \tilde{\alpha}_{21})^T$ as $\tilde{L}_n(\boldsymbol{\theta})$. Theorems \ref{thm2} and \ref{thm3} show that there always exists a sequence $\hat{\boldsymbol{\theta}}_n$, which is a local maximizer of $\tilde{L}_n(\boldsymbol{\theta})$, such that $\hat{\boldsymbol{\theta}}_n$ is consistent and asymptotically normal, regardless of the initial value $(\tilde{\sigma}_1, \tilde{\alpha}_{11}, \tilde{\alpha}_{21})^T$.

\begin{theorem}[Consistency]\label{thm2}
	Assume $\Theta$ is a compact set of $\Theta_s$. Suppose the observations $\{Q_t \}_{t=1}^n$ are generated by a stationary and ergodic model with true parameter $\boldsymbol{\theta}_0$ and $\boldsymbol{\theta}_0$ is in the interior of $\Theta$, then there exists a sequence $\hat{\boldsymbol{\theta}}_n$ of local maximizer of $\tilde{L}_n(\boldsymbol{\theta})$ such that $\hat{\boldsymbol{\theta}}_n \to_p \boldsymbol{\theta}_0$ and $|| \hat{\boldsymbol{\theta}}_n - \boldsymbol{\theta}_0 || \leq \tau_n$, where $\tau_n=O_p(n^{-r})$, $0<r<1/2$. Hence $ \hat{\boldsymbol{\theta}}_n$ is consistency.
\end{theorem}

Theorem \ref{thm2} shows that there exists a sequence $\hat{\boldsymbol{\theta}}_n$ which contains not only consistent cMLE to $\boldsymbol{\theta}_0$ but also local maximizer of $\tilde{L}_n(\boldsymbol{\theta})$. Next, we derive the asymptotic distributions of our estimators $\hat{\boldsymbol{\theta}}_n$ in the following Theorem \ref{thm3}.

\begin{theorem}[Asymptotic normality]\label{thm3}
	Under the conditions in Theorem \ref{thm2}, we have $\sqrt{n} (\hat{\boldsymbol{\theta}}_n-\boldsymbol{\theta}_0) \stackrel{d}{\rightarrow} N(\boldsymbol{0}, \mathbf{M}_0^{-1})$, where $\hat{\boldsymbol{\theta}}_n$ is given in Theorem \ref{thm2} and $\mathbf{M}_0$ is the Fisher Information matrix evaluated at $\boldsymbol{\theta}_0$. Further, the sample variance-covariance matrix of plug-in estimated score functions $\{ \frac{\partial}{\partial \boldsymbol{\theta}} l_t(\hat{\boldsymbol{\theta}}_n)\}_{t=1}^{n}$ is a consistent estimator of $\mathbf{M}_0$.
\end{theorem}

Although the consistency of $\hat{\boldsymbol{\theta}}_n$ and their asymptotic distributions are showed in Theorem \ref{thm2} and Theorem \ref{thm3} respectively, the uniqueness of cMLE remains open due to the complexity brought by $\mu$. Proposition \ref{prop1} provides a segmentary answer to the uniqueness of cMLE.

\begin{proposition}[Asymptotic uniqueness]\label{prop1}
	Denote $V_n=\{\boldsymbol{\theta} \in \Theta | \mu \leq cQ_{n,1} + (1-c) \mu_0 \}$ where $Q_{n,1}=\min_{1\leq t \leq n} Q_t$, under the conditions in Theorem \ref{thm2}, for any fixed $0<c<1$. There exists a sequence of $\hat{\boldsymbol{\theta}}_n=\arg\max_{\boldsymbol{\theta} \in V_n} \tilde{L}_n(\boldsymbol{\theta})$ such that, $\hat{\boldsymbol{\theta}}_n \to_p\boldsymbol{\theta}_0$, $|| \hat{\boldsymbol{\theta}}_n - \boldsymbol{\theta}_0|| \leq \tau_n$ where $\tau_n=O_p(n^{-r})$ with $0<r<1/2$, and $$P(\hat{\boldsymbol{\theta}}_n ~is ~the~ unique~ global~ maximizer~ of~ \tilde{L}_n(\boldsymbol{\theta}) ~over ~V_n) \to 1.$$
\end{proposition}

The proofs of Theorems \ref{thm2} and \ref{thm3} and Proposition \ref{prop1} can be found in the Appendix.

\section{Simulation study}\label{sec: sim}
\subsection{Performance of the conditional maximum likelihood estimator}\label{sec:cmle}
In this section, we study the finite sample performance of the cMLE. We generate data from the AcAF model with the following parameters $(\beta_0, \beta_1, \beta_2, \beta_3, \gamma_0, \gamma_1, \gamma_2, \gamma_3, \delta_0, \delta_1, \delta_2, \delta_3, \mu)^T=(-0.244, 0.787,
0.066, 8.111, 0.23,0.755, 0.417, 7.114, -0.035, 0.907, 0.425,4.861, -0.227)^T$. This set of parameters is obtained from the real data analysis of the S$\&$P 500 daily negative log-returns using the AcAF model. Under this setting, the typical range of $\alpha_{1t}$ is $[3.26, 10.17]$, the typical range of $\alpha_{2t}$ is $[2.68, 26.94]$, and the typical range of $\sigma_t$ is $[0.25, 0.31]$.

We investigate the performance of cMLE with sample sizes $N=1000, 2000, 5000, 10000$. For each sample size, we conduct 100 experiments. The results for parameter estimation are in Table \ref{table1}, including the average of the estimates and the standard deviation from the 100 experiments. From Table \ref{table1}, we can see that both the bias and variance of the cMLE decrease as the sample size $N$ increases, demonstrating the consistency of the cMLE under correct model specification. We find that the performance of cMLE is already satisfactory when $N=5000$.

\begin{table}
	\caption{Numerical results for performance of cMLE with sample sizes 1000, 2000, 5000 and 10000. Mean and S.D. are the sample mean and standard deviation of the cMLE's obtained from 100 simulations.}
	\label{table1}
	\centering
	\begin{tabular}{@{}ccrcrcrcrc@{}}
		\hline
		& &\multicolumn{2}{c}{$N=1000$}&\multicolumn{2}{c}{$N=2000$}
		&\multicolumn{2}{c}{$N=5000$}&\multicolumn{2}{c}{$N=10000$}\\
		\cline{3-10}
		Parameter & True value &
		\multicolumn{1}{c}{Mean} &
		S.D.&
		\multicolumn{1}{c}{Mean} &
		S.D.&
		\multicolumn{1}{c}{Mean} &
		S.D.&
		\multicolumn{1}{c}{Mean} &
		S.D. \\
		\hline
		$\gamma_0$ & 0.230  &0.269 & 0.167 & 0.248 & 0.154 & 0.234 & 0.119 & 0.213& 0.091 \\
		$\gamma_1$  & 0.755  & 0.738& 0.140 & 0.749  & 0.106 & 0.759 & 0.070 &  0.766 &  0.060\\
		$\gamma_2$  & 0.417  & 0.440 & 0.208  & 0.438  & 0.157 &0.427 & 0.088 & 0.428 & 0.075 \\
		$\gamma_3$  & 7.114& 7.507 & 2.805 & 7.589  & 2.706 & 7.385 &1.939 &7.083 & 1.737 \\
		$\delta_0$   & $-$0.035   & $-$0.011 & 0.058  & $-$0.009 &0.053 & $-$0.010 & 0.052 & $-$0.017 &0.047   \\
		$\delta_1$   & 0.907  & 0.886& 0.060  & 0.890  & 0.052 &0.896 &0.036& 0.897&  0.032  \\
		$\delta_2$   & 0.425 & 0.475 & 0.183 & 0.446 & 0.133 & 0.436 & 0.083 & 0.435 & 0.068 \\
		$\delta_3$   & 4.861   & 5.759 & 2.287  & 5.498  & 1.776 &  5.330 & 1.346&  5.134 & 1.061\\
		$\beta_0$   & $-$0.244  & $-$0.227 & 0.090  & $-$0.236 & 0.063 & $-$0.234 & 0.042 & $-$0.235 & 0.032 \\
		$\beta_1$   & 0.787   &0.767 & 0.065 &0.781  & 0.043& 0.782 &  0.022 &0.784 & 0.015  \\
		$\beta_2$   & 0.066  &0.083 & 0.052&0.072  & 0.029& 0.065 &  0.015&0.064& 0.010 \\
		$\beta_3$   & 8.111   &7.348& 3.555 &8.097 & 2.845& 8.216 & 1.968&8.107 & 1.586 \\
		$\mu$   & $-$0.227& $-$0.267  &0.098& $-$0.249&0.083 &$-$0.252& 0.056 &  $-$0.247 &0.039 \\
		\hline
	\end{tabular}
\end{table}

\subsection{Comparison with the autoregressive conditional Fr\'echet model}
\begin{figure}
	\centering
	\includegraphics[width=0.98\textwidth]{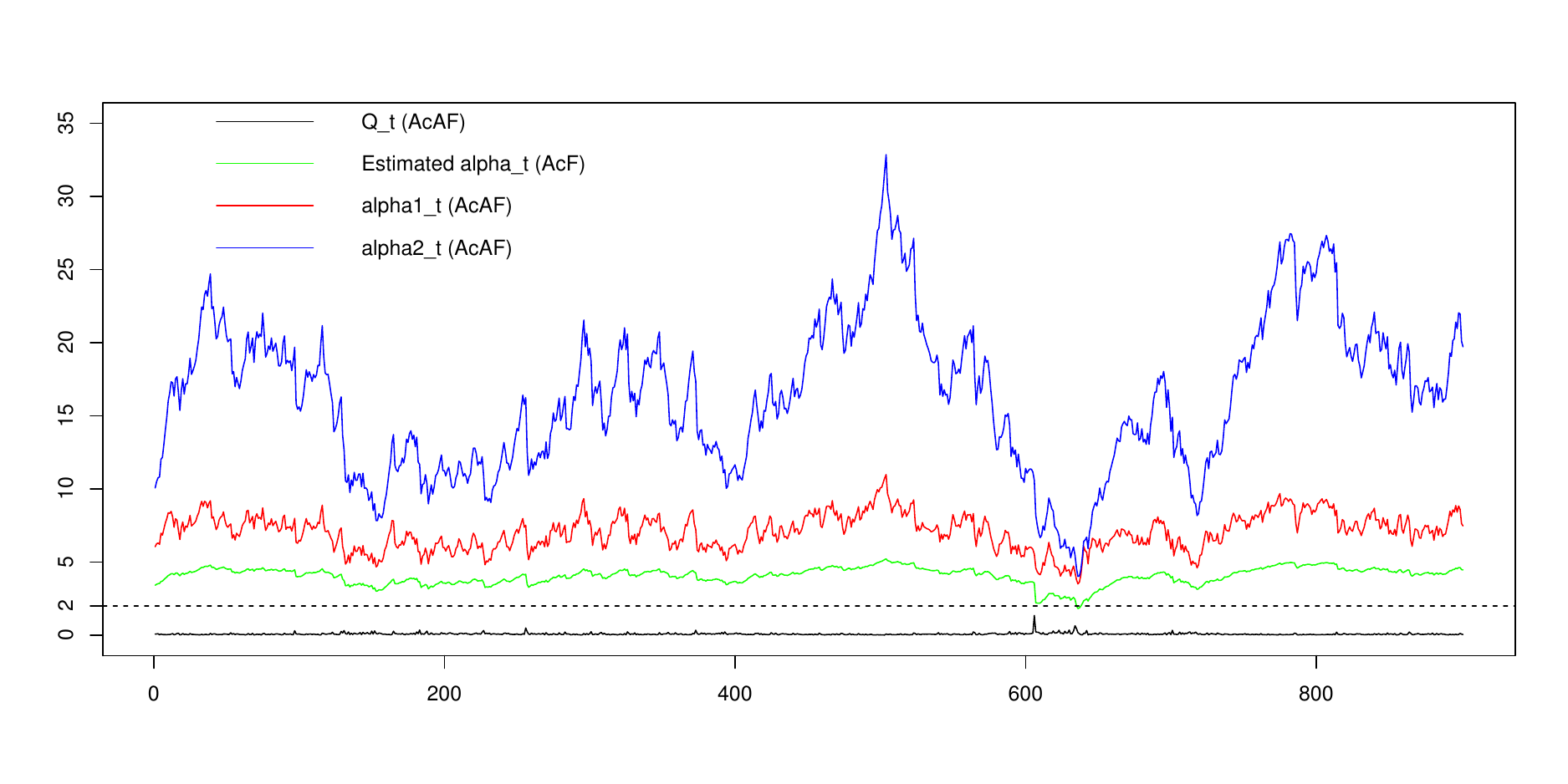}
	\caption{Estimated tail index $\{\hat{\alpha}_{t} \}$ (green) by the AcF model, recovered tail indices $\{\alpha_{1t} \}$ (red), $\{\alpha_{2t} \}$ (blue) by the AcAF model and simulated $\{Q_t\}$ (black) by the AcAF model. All plotted series omit the first 100 data points.}
	\label{fig01}
\end{figure}

In this section, we compare our AcAF model with the autoregressive conditional Fr\'echet (AcF) model. The AcF model is one of the time-varying GEV models that can be used to model time series data of maxima. The AcF model converts $\max(Y_{1t}^{1/\alpha_{1t}},Y_{2t}^{1/\alpha_{2t}})$ in equation (\ref{10}) to $Y_t^{1/\alpha_t}$ and contains the dynamic structures for $\sigma_t$ and $\alpha_t$ same to equations (\ref{11}) and (\ref{12}). More details can be found in \cite{zhao2018modeling}. The way to fulfill the comparison is as follows. First, we simulate data $\{Q_t\}$ with a length of 1000 from the AcAF model by the parameters mentioned in Section \ref{sec:cmle}, and use the simulated data and parameters to recover two tail indices $\{\alpha_{1t}\}$ and $\{\alpha_{2t}\}$ through the evolution structures (\ref{12}) and (\ref{13}). Then we fit the AcF model on the simulated data to estimate its tail index $\hat{\alpha}_t$. The simulated $\{Q_t\}$, recovered tail indices $\{\alpha_{1t}\}$ and $\{\alpha_{2t} \}$ by the AcAF model and the fitted tail index $\{\hat{\alpha}_{t} \}$ by the AcF model are presented in Figure \ref{fig01}.

From Figure \ref{fig01}, we can see that the AcF model seems to give a significantly lower estimation of the tail index than the AcAF model. An under-estimated tail index implies over-estimated tail risk, which in turn may result in higher reserve requirements and other expenses for financial institutions, and in turn lead to reduced liquidity of financial institutions. If the potential loss (risk) that a financial system faces based on the known market information has multiple sources, that is, the loss data is not i.i.d., the tail index estimated by the AcF model is inaccurate and very different from the real tail risk. The recovered tail indices $\{\alpha_{1t}\}$ and $\{\alpha_{2t}\}$ by the AcAF model are all larger than 2, hence the conditional mean and variance of the simulated maxima of maxima $\{Q_t\}$ from the AcAF model always exist. We note that the estimated tail index $\{\hat{\alpha}_{t} \}$ by the AcF model is less than 2 at some points, which means that the variance does not exist.

\begin{figure}
	\centering
	\includegraphics[width=0.98\textwidth]{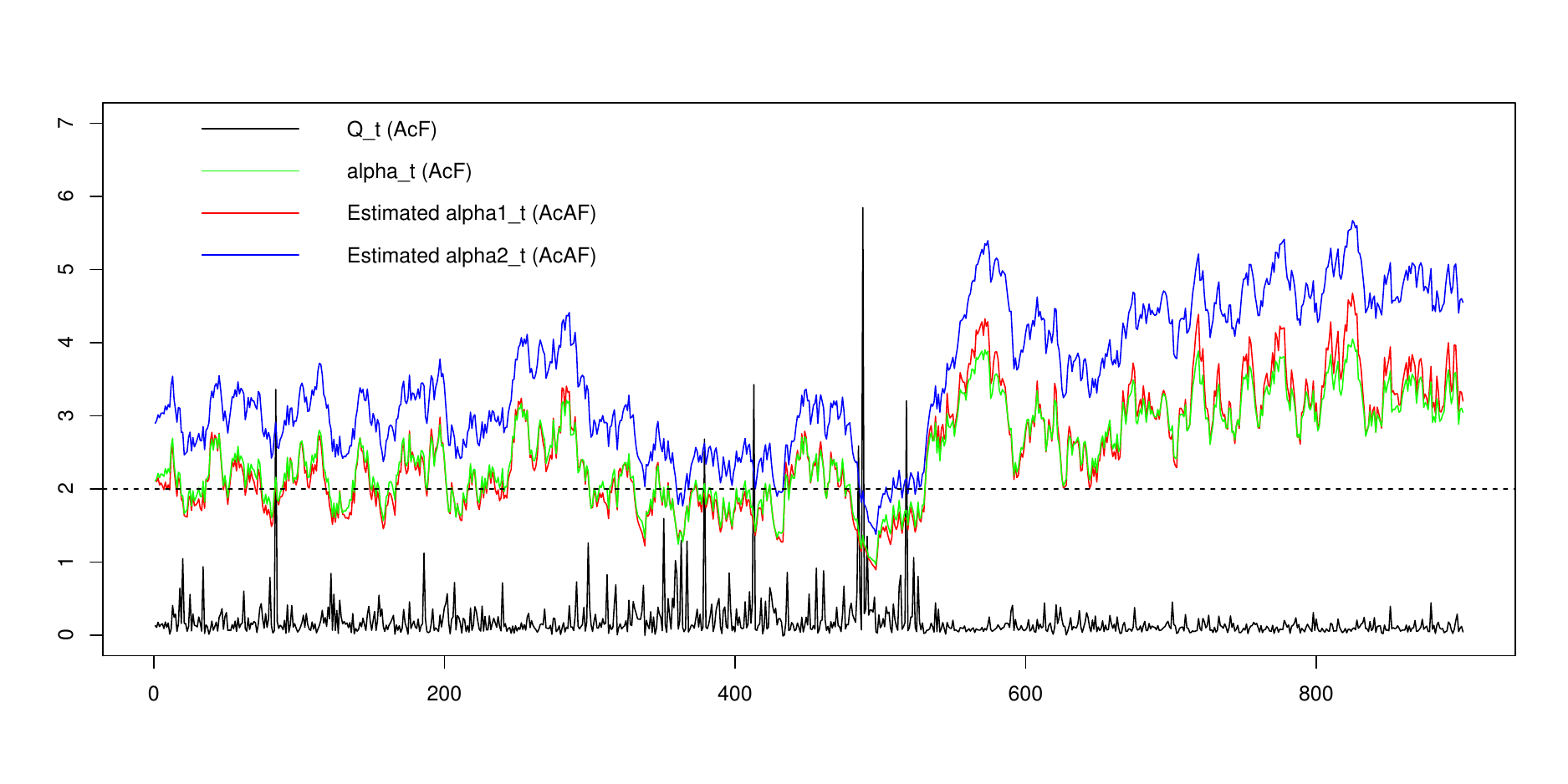}
	\caption{Recovered tail index $\{\alpha_{t} \}$ (green) by the AcF model, estimated tail indices $\{\hat{\alpha}_{1t} \}$ (red), $\{\hat{\alpha}_{2t} \}$ (blue) by the AcAF model and simulated $\{Q_t\}$ (black) by the AcF model. All plotted series omit the first 100 data.}
	\label{fig02}
\end{figure}

We also simulate $\{Q_t\}$ with a length of 1000 by the AcF model with the parameters provided in \cite{zhao2018modeling}, then apply the AcAF model to estimate parameters. Similarly, we recover the tail index $\{\alpha_t\}$ of the simulated data and plot it with the estimated tail indices $\{\hat{\alpha}_{1t} \}$  and $\{\hat{\alpha}_{2t} \}$ of the AcAF model in Figure \ref{fig02}.

As can be seen from Figure \ref{fig02}, the tail index recovered by the AcF model almost coincides with $\{\hat{\alpha}_{1t} \}$ estimated by the AcAF model, i.e., the green line in Figure \ref{fig02} coincides with the red line. In the AcAF model, tail risk is mainly characterized by the dominant sequence and the $\{\hat{\alpha}_{1t}\}$ series captures the dominant information, which indicates that an AcAF model fitting is acceptable for the data generated from an AcF model. And the  conditional mean and variance of the maxima of maxima $\{Q_t\}$ do not always exist.

In summary, using the patterns in Figures \ref{fig01} and \ref{fig02}, we can visually conclude which model, AcF or AcAF, to be used in real data analysis and make inferences.

\subsection{Convergence of maxima of maxima in factor model}
In this section, we conduct numerical experiments to investigate the finite sample behavior of $Q_t$ described in Corollary \ref{coro1}. Specifically, we study the convergence of the marginal distribution of $Q_t$ to its accelerated Fr\'echet limit under a one-time period factor model. To simplify notation, we drop the time index $t$ in this section. We simulate data from the following one-factor linear model,
\begin{equation*}
X_i=\beta_i Z+\sigma_i \max(\epsilon_{1i}, \epsilon_{2i}), ~~i=1,\cdots,p,
\end{equation*}
where $Z \sim N(0,1)$ is the latent factor, $\beta_i$'s are i.i.d. random coefficients generated from a uniform distribution $U(-2,2)$ and $\sigma_i$'s are i.i.d. random variables generated from $\frac{1}{2}U(0, 0.09)+\frac{1}{2}U(0.01, 0.08)$ such that all $\sigma_i$'s are moderate in $(0.005, 0.085)$. This setting roughly matches the pattern of GARCH(1,1) fitted average volatilities of 505 different stocks in S\&P 500.

Random variables $\epsilon_{1i}$'s,  $\epsilon_{2i}$'s are independent and their distributions are in the Fr\'echet domain of attraction. We will select different combinations of $\epsilon_{1i}$ and $\epsilon_{2i}$ according to the relationship between two tail indices. Specifically, we will choose the following three cases to perform our simulation study: (i) $t(3)$ and $t(5)$; (ii) $t(3)$ and Pareto$(1, 3)$; (iii) $t(3)$ and $t(2)$. Here $t(\nu)$ represents $t$-distribution with degree of freedom $\nu$, and Pareto$(x_m, \alpha)$ represents Pareto distribution with c.d.f. $F(x)=1-(x_m/x)^\alpha$ for $x \geq x_m$. This setting corresponds to the three cases in Corollary \ref{coro1}.

\begin{figure}[htbp]
	\centering
	\subfigure[]{
		\includegraphics[width=5cm]{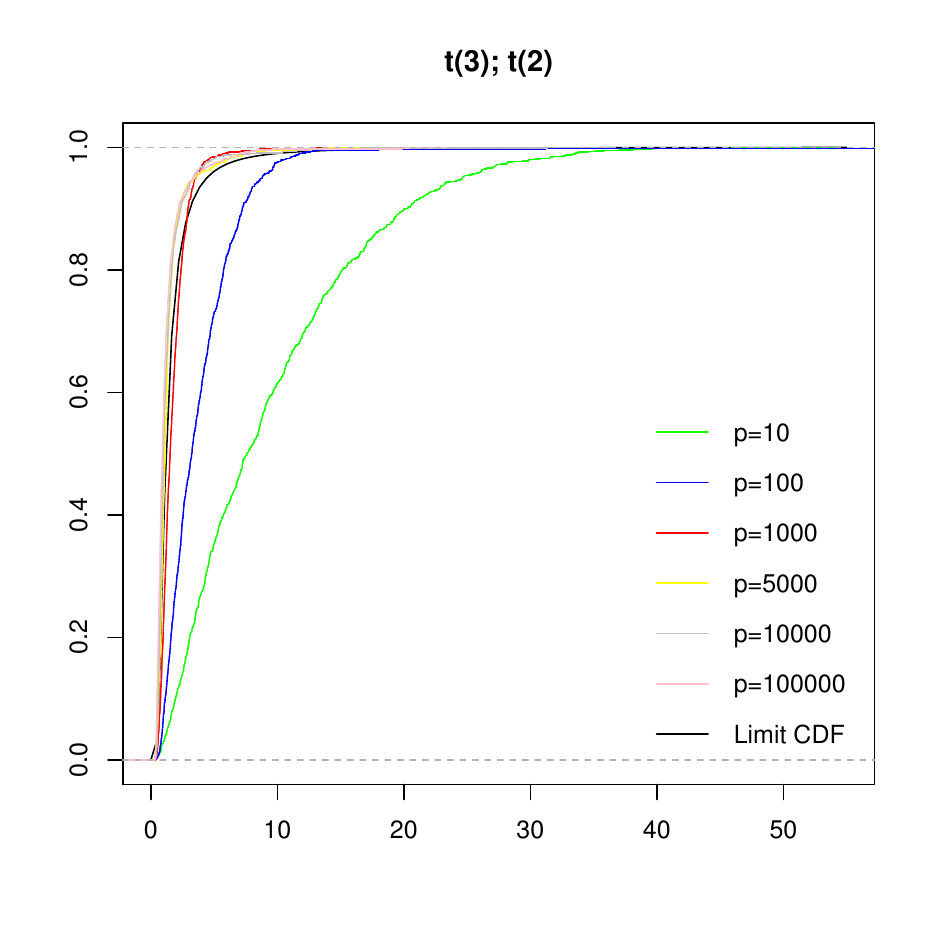}
	}
	\subfigure[]{
		\includegraphics[width=5cm]{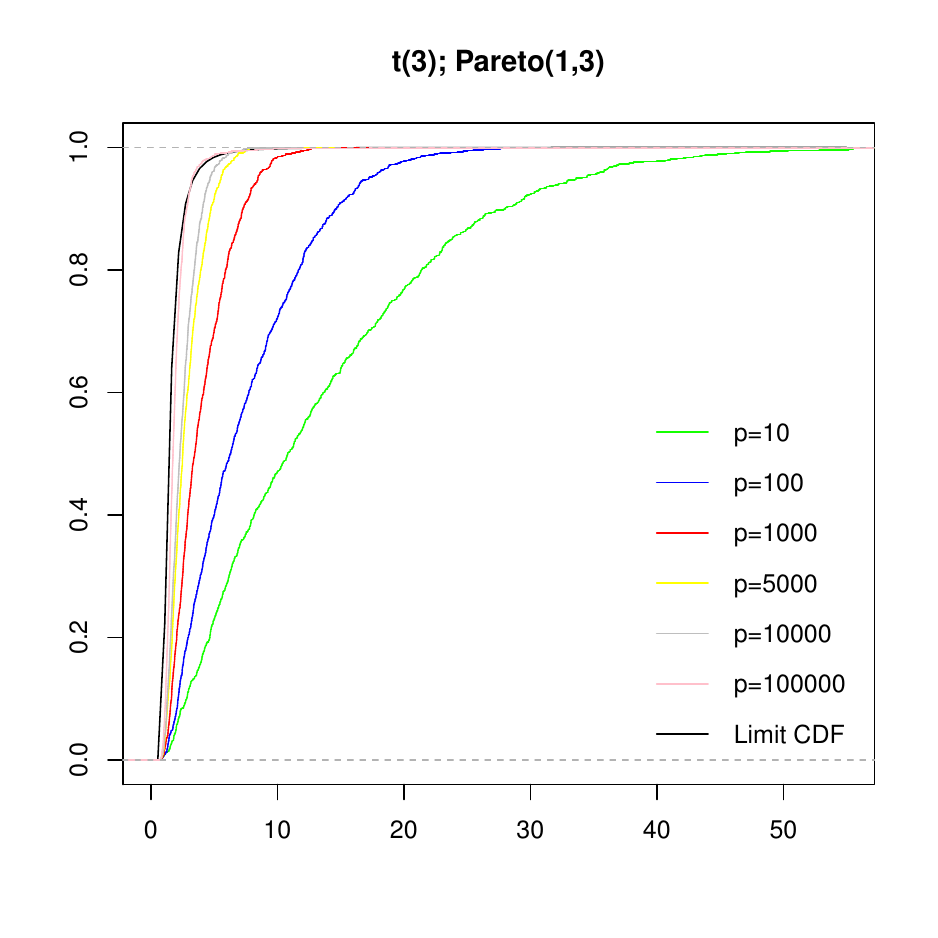}
	}
	\subfigure[]{
		\includegraphics[width=5cm]{0314t3t2-eps-converted-to.pdf}
	}
	\caption{Finite sample empirical distribution of the maxima of maxima Q and its corresponding accelerated Fr\'echet limit, with different combinations of p and distributions of $\left( \epsilon_{1i}, \epsilon_{2i} \right)^T$ in the factor model.}
	\label{fig_factor0}
\end{figure}

We set $Q=\max_{1 \leq i \leq p}X_i$ and compare the finite sample empirical distribution of $Q$ and its corresponding limit stated in Corollary \ref{coro1} under different $\epsilon_{1i}$, $\epsilon_{2i}$ and $p$. For each $(\epsilon_{1i}, \epsilon_{2i}, p)^T$ combination, 1000 sets of i.i.d. $\{X_{i}\}_{i=1}^p$ are generated, resulting in 1000 sampled $Q=\max_{1 \leq i \leq p}X_i$. Figure \ref{fig_factor0} plots the empirical c.d.f. of the normalized $Q$ in Corollary \ref{coro1} along with the corresponding limiting accelerated Fr\'echet distribution. It can be clearly see in Figure \ref{fig_factor0} that as $p$ increases, the empirical distribution of $Q$ approaches its accelerated Fr\'echet limit. A larger tail index requires a larger $p$ for accurate approximation.
.

\section{Real data applications}\label{sec: real}
In this section, we present three real data applications of the AcAF model, one on the cross-sectional maxima of negative log-returns of stocks in S\&P 500, one on the intra-day maxima of negative log-returns from high-frequency stock trading, and the other on the intra-day maxima of negative log-returns from high-frequency bitcoin trading. Notice that the maxima here is defined by the maxima of two maxima, but it is equivalent to taking maxima across all stocks' negative log-returns or high-frequency negative log-returns, so we will use these two concepts ``maxima'' and ``maxima of maxima'' interchangeably. In all three cases, the AcAF model shows its superiority over the traditional autoregressive tail index models for modeling the endopathic and exopathic competing tail risks in the financial market.

\subsection{Cross-sectional maxima of the daily negative log-returns of stocks in S\&P 500}
In this section, we consider the cross-sectional maxima of the daily negative log-returns (i.e., daily losses) of component stocks in the S\&P 500 Index. S\&P 500 Index is an American stock market index based on the market capitalizations of 505 large companies, which is among the most commonly followed equity indices and the best representations of the U.S. stock market. To better manage the risk, mutual funds and banks must understand the cross-sectional tail risk of S\&P 500. Our data contains the daily closing prices of 505 components of S\&P 500 and is downloaded from Yahoo Finance with the time range January 1, 2005, to August 31, 2020.

\begin{figure}
	\centering
	\includegraphics[width=0.98\textwidth]{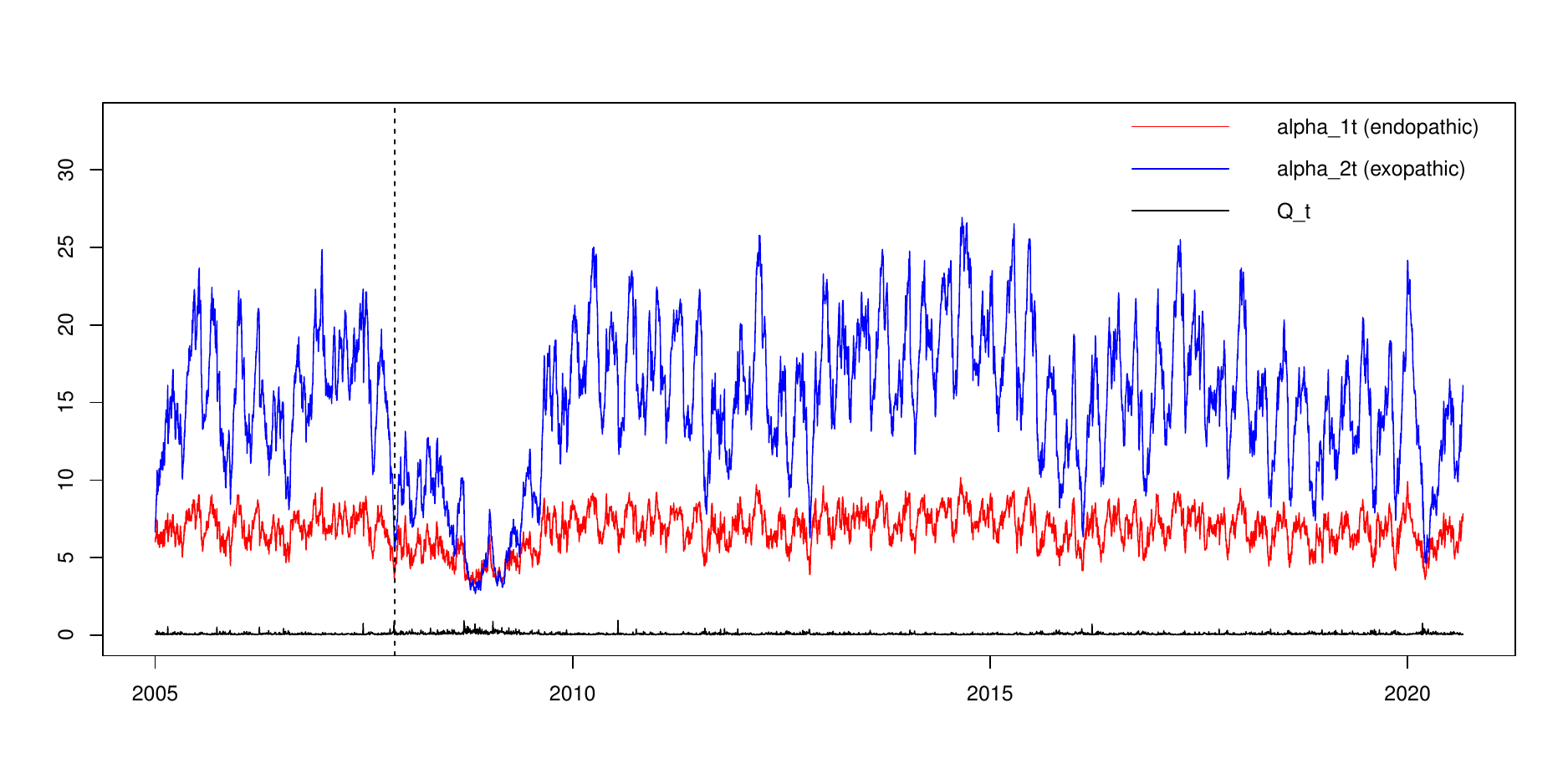}
	\caption{Estimated tail indices $\{\hat{\alpha}_{1t} \}$ (red), $\{\hat{\alpha}_{2t} \}$ (blue) and cross-sectional maximum daily negative log-returns $\{Q_t\}$ (black) of S\&P500 Index from January 3, 2005 to August 31, 2020. The sample variances of $\gamma_2 \exp(-\gamma_3 Q_{t})$ and $\delta_2 \exp(-\delta_3 Q_{t})$ are 0.00591 and 0.00484, respectively.}
	\label{fig1}
\end{figure}

We present the modeling results for S\&P 500 in detail. For each trading day $t$, we calculate the daily negative log-returns of each component stock in S\&P 500 and then calculate the daily cross-sectional maxima $Q_t=\max_{1 \leq i \leq 505}r_{it}$, where $r_{it}$ is the daily negative log-return for stock $i$. The time series $\{Q_t\}$ contains 3934 observations and is shown in the bottom panel of Figure \ref{fig1}.

The estimation results of our model are summarized in Table \ref{table9}. From the results we can see: the estimated autoregressive parameter values of $\hat{\gamma}_1$ and $\hat{\delta}_1$ for the tail indices $\{\alpha_{1t}\}$ and $\{\alpha_{2t}\}$ are both close to 1, which suggests a strong persistence of the tail risk processes. The estimated tail indices $\{\hat{\alpha}_{1t}\}$ and  $\{\hat{\alpha}_{2t}\}$ are plotted in Figure \ref{fig1}.  The range of estimated tail index for endopathic risk is roughly within $[3.26, 10.17]$, while the one for exopathic risk is $[2.68, 29.94]$.  Obviously, when the extreme events appear, two tail indices both tend to decrease, reflecting an increase in risk. Moreover, we can see that exopathic risks are more volatile than endopathic risks, especially when extreme events occur. Under normal market conditions, endopathic risks dominate the stock market price fluctuations, while under turbulent market conditions, exopathic risks dominate. This phenomenon shows that $\alpha_{1t}$ and $\alpha_{2t}$ are useful measures of the endopathic and exopathic tail risks, respectively, and our model has a strong ability to capture information of financial crisis, i.e., exopathic risks dominate endopathic risks.

Like what is found in \cite{zhao2018modeling}, the endopathic and exopathic tail indices of S\&P 500 experienced sudden downside movement around the end of 2007, which reached their lowest level for the past several years, and the exopathic tail risk index dropped sharply, breaking through the endopathic risk, taking a dominant role. As is said in \cite{zhao2018modeling}, this unusual movement can be viewed as a warning signal of the 2008 financial crisis. Based on Figure \ref{fig1}, we see that the patterns of endopathic risks and exopathic risks can better describe and predict a potential crisis.

A Fr\'echet type random variable has its $k$-th moment if and only if $\alpha >k$. It is also noted that all $\hat{\alpha}_{1t}$ and $\hat{\alpha}_{2t}$'s are larger than 2, hence the conditional mean and variance of the cross-sectional maxima always exist, which agrees with the existing literature, e.g., \cite{hansen1994autoregressive}, and is contrary to some literature findings of the tail index being less than 2 due to a single type of Fr\'{e}chet distribution specification.

\begin{figure}
	\centering
	\includegraphics[width=0.98\textwidth]{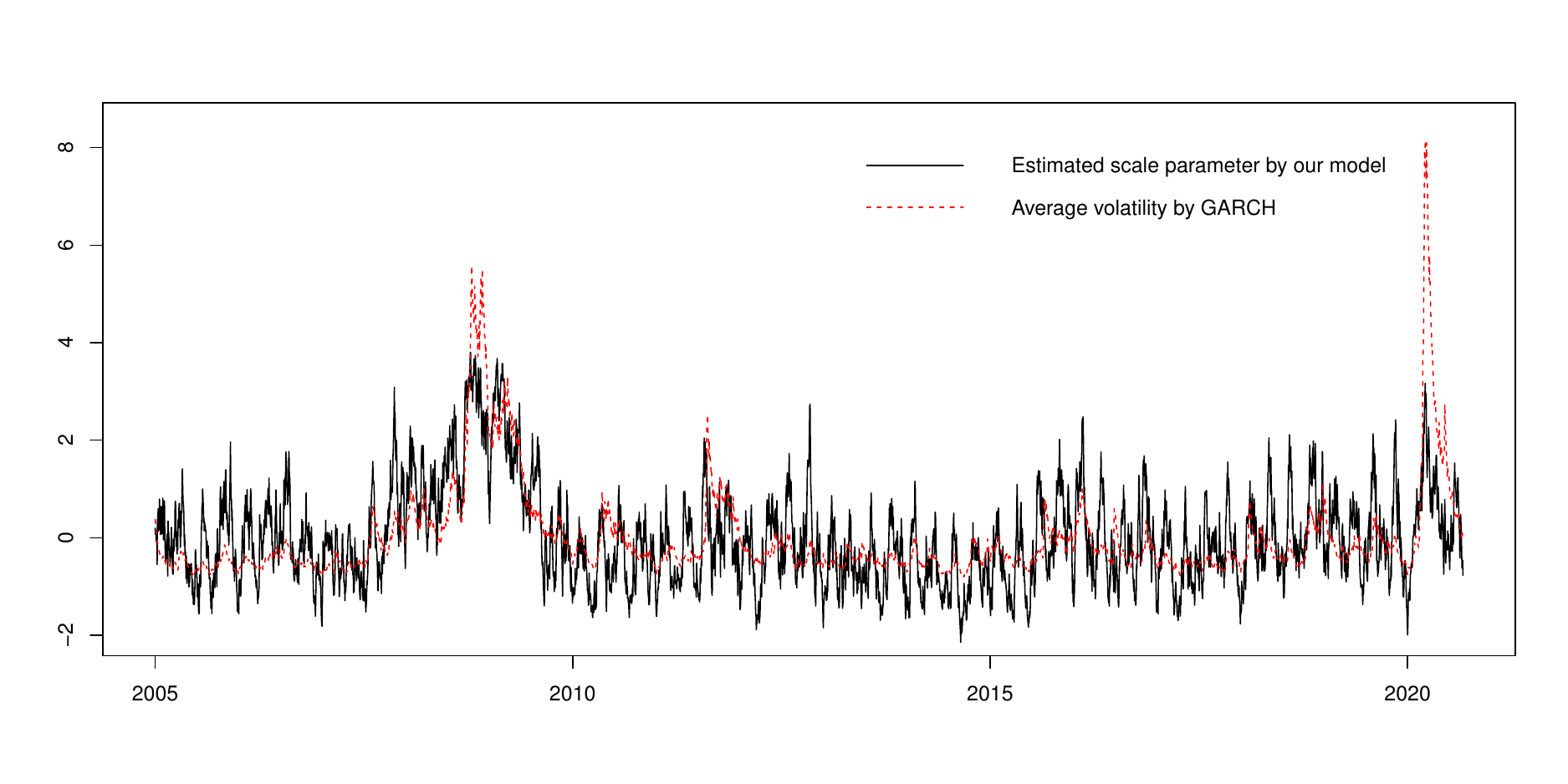}
	\caption{Estimated scale parameter series $\{\hat{\sigma}_t\}$ of S\&P500 Index (black) by our model v.s. estimated average volatility series by GARCH (red) from January 3, 2005 to August 31, 2020. Both series are standardized to be zero mean and unit variance for comparison.}
	\label{fig2}
\end{figure}

The estimated scale parameter $\{\hat{\sigma}_t\}$ by our model is shown in Figure \ref{fig2}. For comparison, we also fit a GARCH(1,1) model for each component stock in S\&P 500 and plot the daily average volatility given by the GARCH model across the 505 stocks in Figure \ref{fig2}. These two series move very closely with each other, with an overall correlation of 0.65. It suggests that our model's dynamic scale parameter $\sigma_t$ is an accurate measure of market volatility. Our results are consistent with the ideas in \cite{danielsson2003endogenous}, indicating the stock market is subject
to both types of risk. The greatest damage which the stock market are subjected to is done from the risk of the endopathic kind, especially in normal market conditions.

\begin{table*}
	\caption{Estimated parameters and standard deviations for S\&P500 from January 1, 2005 to August 31, 2020.}
	\label{table9}
	\centering
	\begin{tabular}{@{}lrrrrrrrr@{}}
		\hline
		\multicolumn{1}{c}{}
		& \multicolumn{1}{c}{$\gamma_0$} & \multicolumn{1}{c}{$\gamma_1$}
		& \multicolumn{1}{c}{$\gamma_2$} & \multicolumn{1}{c}{$\gamma_3$}
		& \multicolumn{1}{c}{$\delta_0$} & \multicolumn{1}{c}{$\delta_1$}
		& \multicolumn{1}{c}{$\delta_2$} & \multicolumn{1}{c}{$\delta_3$}\\
		\hline
		Estimates &0.230 &0.755 &0.417& 7.114 & $-$0.035 & 0.907 & 0.425 &4.861 \\
		S.D. & 0.149 &0.082&0.148&5.007&0.064&0.006&0.003&1.445 \\
		\hline
		\multicolumn{1}{c}{} & \multicolumn{1}{c}{$\beta_0$} & \multicolumn{1}{c}{$\beta_1$}
		& \multicolumn{1}{c}{$\beta_2$} & \multicolumn{1}{c}{$\beta_3$}
		& \multicolumn{1}{c}{$\mu$}\\
		\hline
		Estimates &$-$0.244 & 0.787 & 0.066 & 8.111 & $-$0.227 \\
		S.D.&0.058&0.027&0.021&2.473&0.075\\
		\hline
	\end{tabular}
\end{table*}

\subsection{Intra-day maxima of 5-min negative log-returns of GE stock}\label{section5.2}
In this section, we consider modeling intra-day maxima of 5-minute negative log-returns of GE stock. We collect the historical 1-minute intra-day GE stock price from January 1, 2008, to June 7, 2013. Then we convert this time series into GE stock prices with time intervals of 5-minute. The 5-minute negative log-returns $\{r_{i,t}\}_{i=1}^{p}$ are obtained and intra-day maxima $Q_t$ are calculated. The total length of series $\{Q_t\}$ is 1356.

We fit the AcAF model to the intra-day maxima 5-minute negative log-returns series. Estimated parameters and their standard deviations are shown in Table \ref{table12}. The estimated autoregressive parameters $\hat{\beta}_1$ for $\{\sigma_t\}$ and $\hat{\delta}_1$ for $\{\alpha_{2t}\}$ are close to 1, showing strong persistence of the scale $\{\sigma_t\}$ and exopathic tail risk index $\{\alpha_{2t}\}$ series; while the autoregressive parameter $\hat{\gamma}_1$ for $\{\alpha_{1t}\}$ is 0.303, indicating a less persistence of endopathic tail risk index $\{\alpha_{1t}\}$ series.

\begin{figure}
	\centering
	\includegraphics[width=0.98\textwidth]{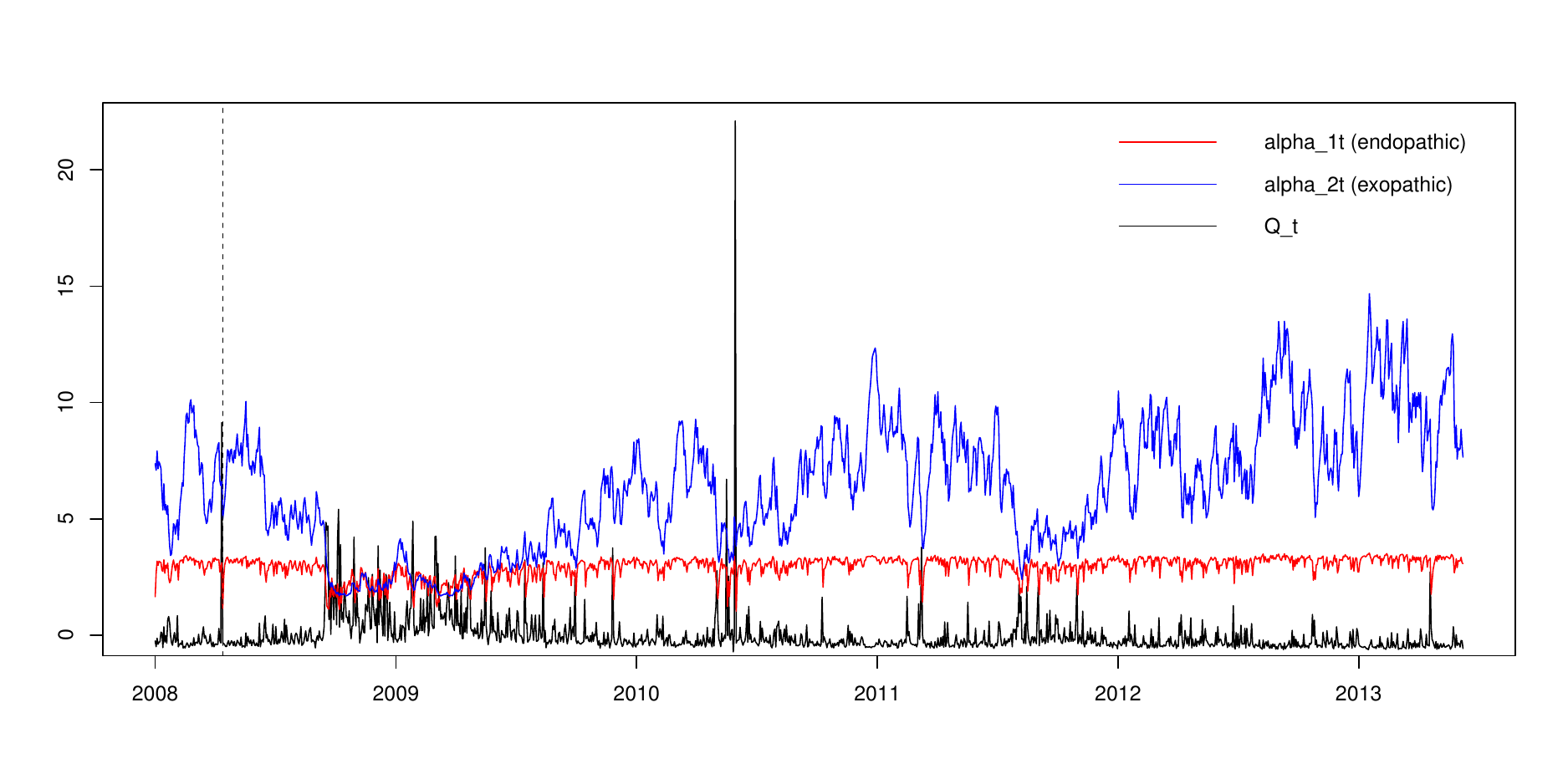}
	\caption{Estimated tail indices $\{\hat{\alpha}_{1t} \}$ (red), $\{\hat{\alpha}_{2t} \}$ (blue) and intra-day maximum of 5-minute negative log-returns $\{Q_t\}$ (black; normalized) for GE stock from January 1, 2008 to June 7, 2013. The sample variances of $\gamma_2 \exp(-\gamma_3 Q_{t})$ and $\delta_2 \exp(-\delta_3 Q_{t})$ are 0.03182 and 0.0143, respectively.}
	\label{fig3}
\end{figure}

The estimated tail indices $\{\hat{\alpha}_{1t}\}$ and  $\{\hat{\alpha}_{2t}\}$ are plotted in Figure \ref{fig3}.  The range of the estimated tail index for endopathic risk is roughly within $[1.74, 5]$, while the one for exopathic risk is $[1.26, 16.15]$.  Obviously, when the extreme events appear, two tail indices both tend to decrease, reflecting an increase in risk. Moreover, we can see that exopathic risks are more volatile than endopathic risks, especially when extreme values occur. Under normal market conditions, endopathic risks dominate the stock market price fluctuations, while under turbulent market conditions, exopathic risks dominate. This phenomenon is interesting and consistent with the market trading behavior, i.e., the market risks are more dominated by endopathic risks, while the exopathic risks caused by the sentiments of investors can be a driving force of large market variations in high-frequency trading. Figure \ref{fig3} also shows that on April 14, 2008, the endopathic tail risk index and exopathic tail risk index plummeted, reaching their lowest values since 2008. This can be regarded as an early warning of the financial crisis that began from September 2008. The estimated scale parameter $\{\hat{\sigma}_t\}$ is showed in Figure \ref{fig4}.

\begin{figure}
	\centering
	\includegraphics[width=0.98\textwidth]{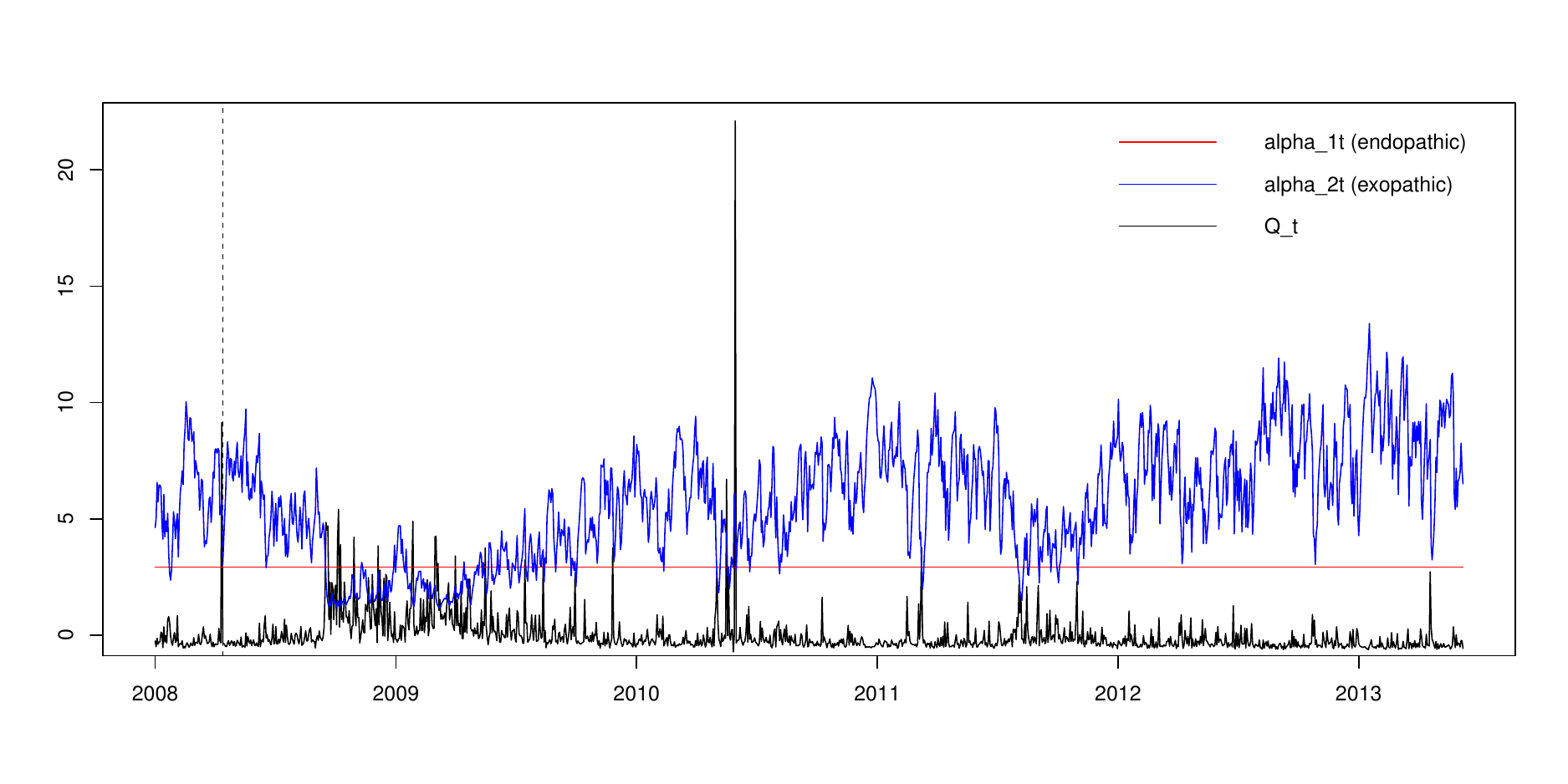}
	\caption{Estimated tail indices $\{\hat{\alpha}_{1t} \}$ (red), $\{\hat{\alpha}_{2t} \}$ (blue) and intra-day maximum of 5-minute negative log-returns $\{Q_t\}$ (black; normalized) of the simplified model for GE stock from January 1, 2008 to June 7, 2013.}
	\label{fig3c}
\end{figure}

\begin{figure}
	\centering
	\includegraphics[width=0.98\textwidth]{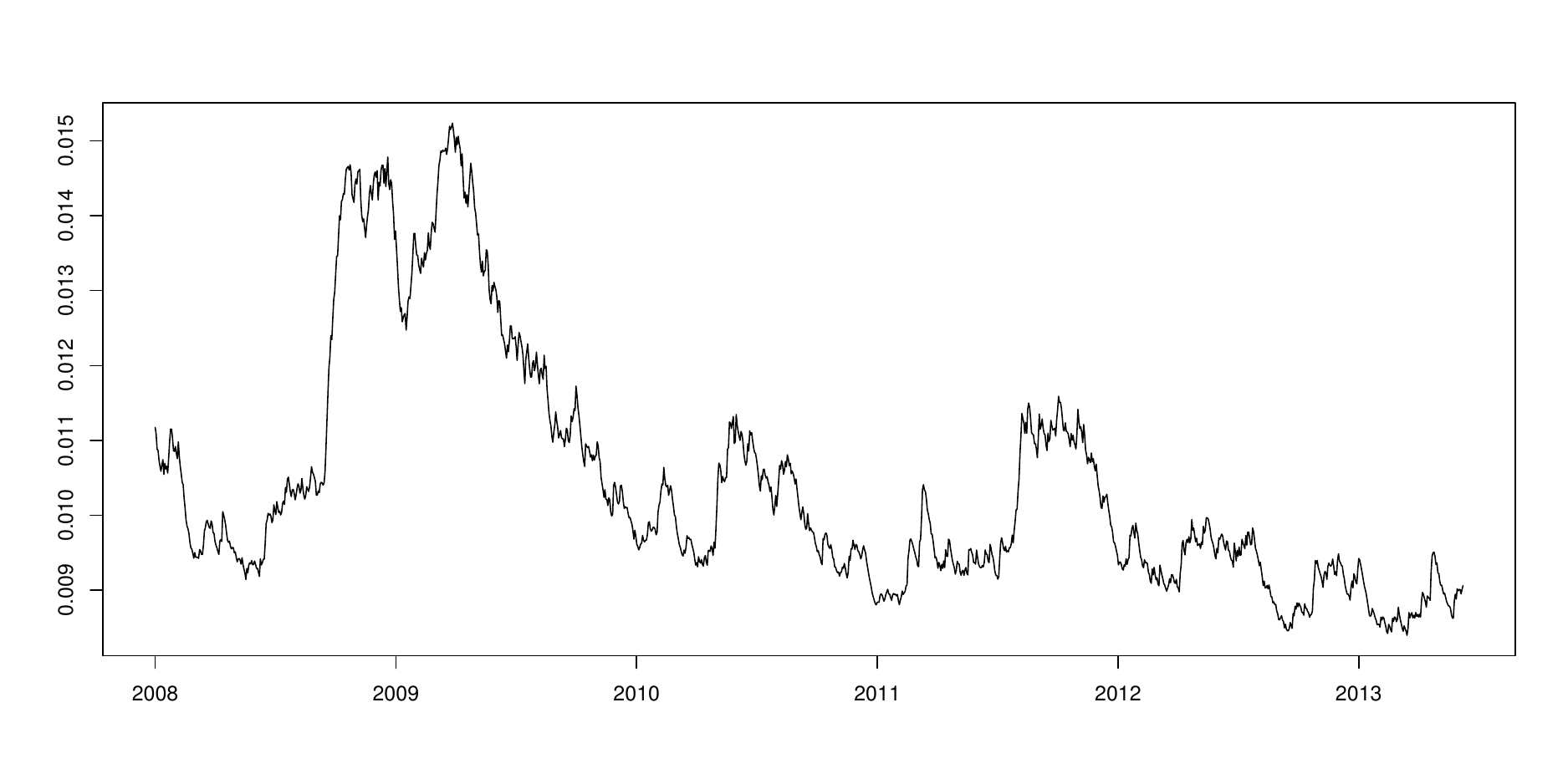}
	\caption{Estimated scale parameter $\{\hat{\sigma}_t\}$ of 5-minute GE stock from January 1, 2008 to June 7, 2013.}
	\label{fig4}
\end{figure}

From Table \ref{table12}, we can see that the estimated standard deviations associated with $\hat{\gamma}_0$, $\hat{\gamma}_1$, $\hat{\gamma}_3$ and $\hat{\delta}_0$ are relatively large, and in Figure \ref{fig3}, compared to the estimated $\{\hat{\alpha}_{2t}\}$, the estimated $\{\hat{\alpha}_{1t}\}$ behaves like a constant except during the 2007-2009 financial crisis period. Following Remark \ref{remark3}, we set $\alpha_{1t}$ as static (i.e., in (\ref{12}) we have $\log \alpha_{1t}=\gamma'$),  which is a simplified model. Performance of cMLE of the simplified model for intra-day maxima of 5-min negative log-returns of GE stock is shown in Table \ref{table12-1}. We can see that all parameters in the simplified model are significant except $\delta_0$. The estimated $\{\hat{\alpha}_{2t}\}$ together with the constant series $\{\alpha_{1t}\}$ are plotted in Figure \ref{fig3c}. We can see that the estimated $\{\hat{\alpha}_{2t}\}$s in Figures \ref{fig3} and \ref{fig3c} are very similar, and we calculated their correlation to be 0.9369. These observations clearly point out the consistency of Definition \ref{riskdecoup} and the interpretations. From both plots, we see that during normal trading days,  the systemic risks faced by large companies such as GE are mainly from internal (endopathic) risks, and during clustered extreme events, e.g., financial crisis, the systemic risks were driven by the exopathic risks.

\begin{table*}
	\caption{Estimated parameters and standard deviations for intra-day maxima of 5-minute negative log-returns of GE stock from January 1, 2008 to June 7, 2013.}
	\label{table12}
	\centering
	\begin{tabular}{@{}lrrrrrrrr@{}}
		\hline
		\multicolumn{1}{c}{}
		& \multicolumn{1}{c}{$\gamma_0$} & \multicolumn{1}{c}{$\gamma_1$}
		& \multicolumn{1}{c}{$\gamma_2$} & \multicolumn{1}{c}{$\gamma_3$}
		& \multicolumn{1}{c}{$\delta_0$} & \multicolumn{1}{c}{$\delta_1$}
		& \multicolumn{1}{c}{$\delta_2$} & \multicolumn{1}{c}{$\delta_3$}\\
		\hline
		Estimates & 0.378&  0.303  & 0.829 & 81.88 & $-$0.160 & 0.842  & 0.670 & 41.49 \\
		S.D. & 0.384&0.306&0.331&65.51&0.106&0.024&0.016&15.32\\
		\hline
		\multicolumn{1}{c}{} & \multicolumn{1}{c}{$\beta_0$} & \multicolumn{1}{c}{$\beta_1$}
		& \multicolumn{1}{c}{$\beta_2$} & \multicolumn{1}{c}{$\beta_3$}
		& \multicolumn{1}{c}{$\mu$}\\
		\hline
		Estimates & $-$0.240 & 0.939& 0.063 & 83.30 & $-$0.007\\
		S.D.& 0.029 &0.006&0.016&24.31&0.003\\
		\hline
	\end{tabular}
\end{table*}

\begin{table*}
	\caption{Estimated parameters and standard deviations of the simplified model for intra-day maxima of 5-minute negative log-returns of GE stock from January 1, 2008 to June 7, 2013.}
	\label{table12-1}
	\centering
	\begin{tabular}{@{}lrrrrrrrrrr@{}}
		\hline
		\multicolumn{1}{c}{}
		& \multicolumn{1}{c}{$\gamma'$}
		& \multicolumn{1}{c}{$\delta_0$} & \multicolumn{1}{c}{$\delta_1$}
		& \multicolumn{1}{c}{$\delta_2$} & \multicolumn{1}{c}{$\delta_3$}
		& \multicolumn{1}{c}{$\beta_0$} & \multicolumn{1}{c}{$\beta_1$}
		& \multicolumn{1}{c}{$\beta_2$} & \multicolumn{1}{c}{$\beta_3$}
		& \multicolumn{1}{c}{$\mu$}\\
		\hline
		Estimates & 1.074 & $-$0.007   & 0.632 & 1.141 & 99.27&  $-$0.261  & 0.936 & $-$0.098 & 83.08&  $-$0.004\\
		S.D. & 0.191&0.280&0.175&0.200&39.94&0.028&0.006&0.019&20.98&0.001\\
		\hline
	\end{tabular}
\end{table*}


We note that both the maxima of cross-sectional negative log-returns of stocks from S\&P 500 and the maxima of intra-day negative log-returns of high-frequency stock trading lead to similar observations of the 2007-2009 financial crisis. This phenomenon reveals that the AcAF model is robust in describing and predicting market downturn periods.

\subsection{Intra-day maxima of 5-min negative log-returns for BTC/USD exchange rate}
In this section, we consider modeling intra-day maxima of 5-minute negative log-returns of Bitcoin trading. We convert 1-minute BTC/USD exchange rates to 5-minute frequency time series and obtain daily maxima of negative log-returns $Q_t$. The exchange rate series we employ is available on Kaggle and includes observations from October 8, 2015, to April 9, 2020. The length of the series $\{Q_t\}$ is 1609.

We fit the model to the intra-day maxima 5-minute negative log-returns series.  Estimated parameters and their standard deviations are shown in Table \ref{table17}. The estimated autoregressive parameters $\hat{\beta}_1$ for $\{\sigma_t\}$ and $\hat{\delta}_1$ for $\{\alpha_{2t}\}$ are close to 1, showing strong persistence of the scale $\{\sigma_t\}$ and exopathic tail risk index $\{\alpha_{2t}\}$ series; while the autoregressive parameter $\hat{\gamma}_1$ for $\{\alpha_{1t}\}$ is 0.416, indicating a less persistence of exopathic tail risk index $\{\alpha_{1t}\}$ series.

\begin{figure}
	\centering
	\includegraphics[width=0.98\textwidth]{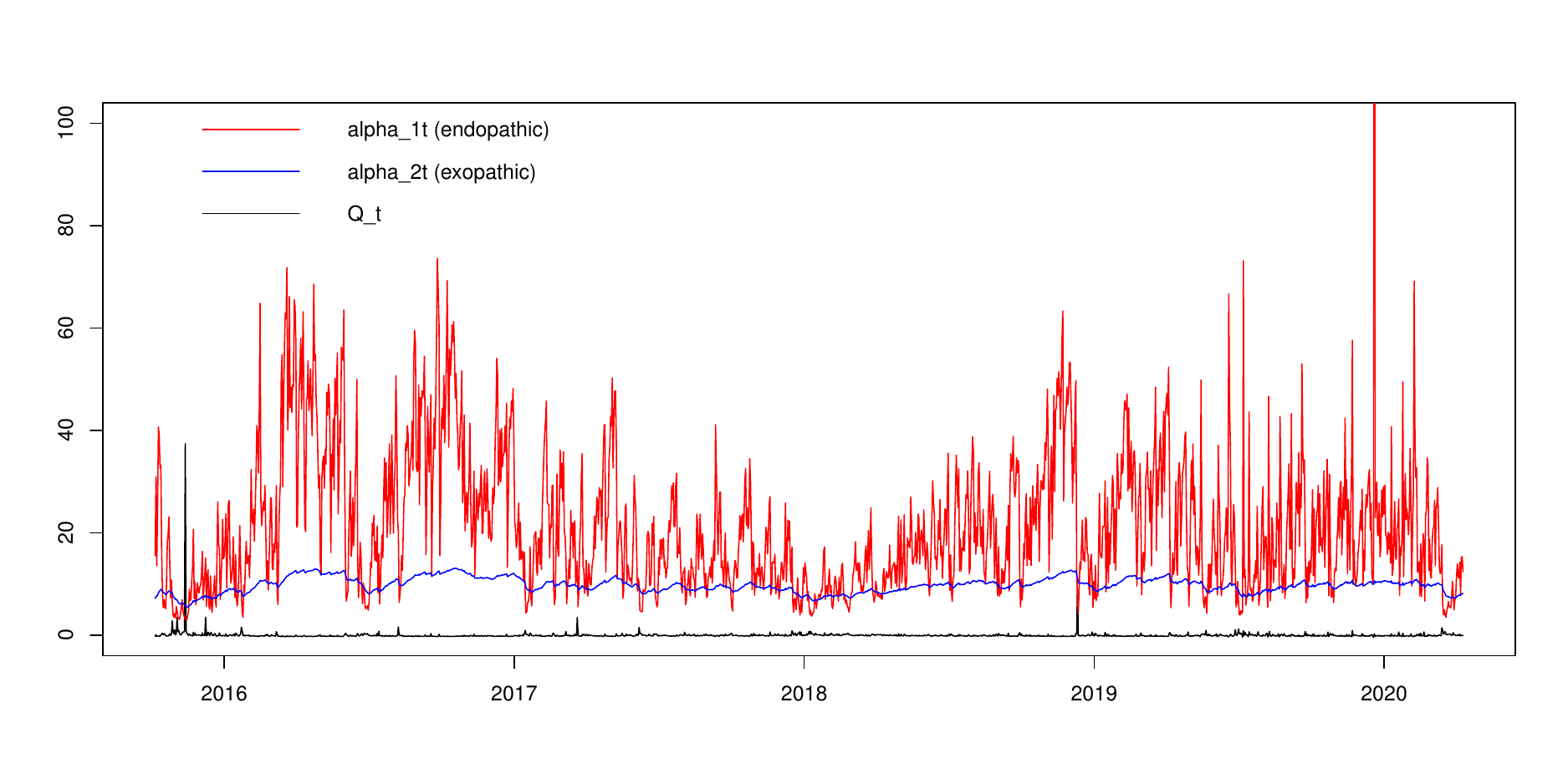}
	\caption{Estimated tail indices $\{\hat{\alpha}_{1t} \}$ (red), $\{\hat{\alpha}_{2t} \}$ (blue) and intra-day maxima of 5-minute negative log-returns $\{Q_t\}$ (black; normalized) from October 8, 2015 to April 9, 2020 for BTC/USD data. The sample variances of $\gamma_2 \exp(-\gamma_3 Q_{t})$ and $\delta_2 \exp(-\delta_3 Q_{t})$ are 0.22677 and 0.00051, respectively.}
	\label{fig9}
\end{figure}

The estimated tail indices $\{\hat{\alpha}_{1t}\}$ and  $\{\hat{\alpha}_{2t}\}$ are plotted in Figure \ref{fig9}.  The range of estimated tail index for endopathic risk is $[2.97, 206.97]$, while the one for exopathic risk is $[5.47, 13.06]$. 
Obviously, when the extreme events appear, two tail indices both tend to decrease, reflecting an increase in risk. Moreover, we can see that endopathic risks are more volatile than exopathic risks, especially when extreme events occur. Exopathic risks dominate the cryptocurrency market price fluctuations under normal market conditions, while under turbulent market conditions, endopathic risks dominate. This is consistent with our empirical understanding of the market. The internal transaction risk of the Bitcoin market leads to a large range of price changes, and external shocks have a relatively small impact on Bitcoin trading. However, for the stock market, the risks brought by its internal trading are relatively stable, and the stock price changes are more susceptible to the impact of external shocks.

\begin{table*}
	\caption{Estimated parameters and standard deviations for intra-day maxima of 5-minute negative log-returns of BTC/USD from October 8, 2015 to April 9, 2020.}
	\label{table17}
	\centering
	\begin{tabular}{@{}lrrrrrrrr@{}}
		\hline
		\multicolumn{1}{c}{}
		& \multicolumn{1}{c}{$\gamma_0$} & \multicolumn{1}{c}{$\gamma_1$}
		& \multicolumn{1}{c}{$\gamma_2$} & \multicolumn{1}{c}{$\gamma_3$}
		& \multicolumn{1}{c}{$\delta_0$} & \multicolumn{1}{c}{$\delta_1$}
		& \multicolumn{1}{c}{$\delta_2$} & \multicolumn{1}{c}{$\delta_3$}\\
		\hline
		Estimates &0.598& 0.416  & 2.004 & 68.573 & 0.100&  0.920  & 0.118  & 35.195\\
		S.D.&0.158&0.055&0.173&12.21&0.054&0.004&0.0005&17.81\\
		\hline
		\multicolumn{1}{c}{} & \multicolumn{1}{c}{$\beta_0$} & \multicolumn{1}{c}{$\beta_1$}
		& \multicolumn{1}{c}{$\beta_2$} & \multicolumn{1}{c}{$\beta_3$}
		& \multicolumn{1}{c}{$\mu$}\\
		\hline
		Estimates &$-$0.470& 0.829 & 0.035 & 63.50 & $-$0.054  \\
		S.D. &0.008&0.029&0.008&15.16&0.009\\
		\hline
	\end{tabular}
\end{table*}

\begin{figure}
	\centering
	\includegraphics[width=0.98\textwidth]{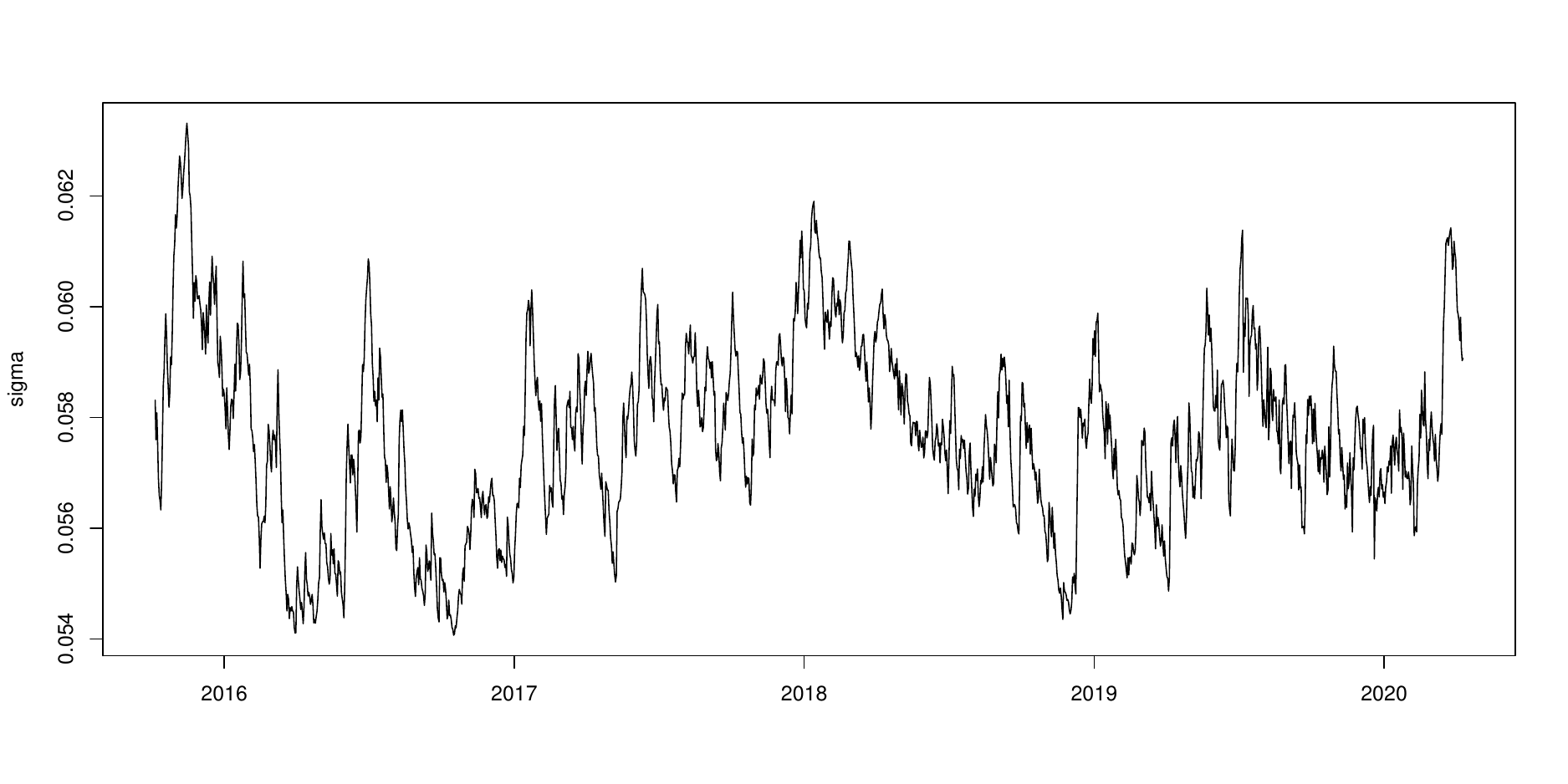}
	\caption{Estimated scale parameter $\{\hat{\sigma}_{t} \}$ from October 8, 2015 to April 9, 2020 for BTC/USD data.}
	\label{fig10}
\end{figure}

The estimated scale parameter $\{\hat{\sigma}_t\}$ is showed in Figure \ref{fig10}.
Comparing Figures \ref{fig4} and \ref{fig10}, we see that the scale parameters in maxima of maxima in Bitcoin returns are 5-10 times larger than those in GE stock price changes, which shows a clear pattern that Bitcoin returns are much more volatile than GE returns in terms of high-frequency trading. The competing patterns of the endopathic risks and the exopathic risks from the stock markets are different from those in the Bitcoin markets, i.e., they have reversed relationship.

\section{Conclusion}\label{sec:con}
This paper develops a new autoregressive conditional accelerated Fr\'echet (AcAF) model for decoupling systemic financial risk into endopathic and exopathic competing risks. We model the worst market returns using maxima of maxima in financial time series, which provides a new angle to identify systemic risk patterns and their impacts in financial markets. The probabilistic properties of stationarity and ergodicity of the AcAF model are investigated. We implement the cMLE for the AcAF model, and the estimators' consistency and asymptotic properties are established. Simulation study shows the AcAF model's superior performance to the existing dynamic GEV models for heterogeneous data and the efficiency of the proposed estimators. The real data examples illustrate its potential broad use in financial risk management and systemic risk monitoring. It provides a clear risk pattern of market risks and the causes of the financial crisis.

The AcAF model can be extended to many other aspects. One potential extension is to assume a dynamic structure for the location parameter $\mu$. Another future direction is to extend two risk sources to multiple sources of risk with the construction of a flexible multivariate dynamic tail risk model.

The AcAF model can be applied to diversified areas as long as decoupling systemic risks into competing endopathic risks and exopathic risks is concerned. These areas include systemic risks in social, political, economic, financial, market, regional, global, environmental, transportation, epidemiological, material, chemical, and physical systems.

\bibliographystyle{apalike}
\bibliography{referencelist}      

\end{document}